\documentclass[journal,twoside]{IEEEtran}
\pdfoutput=1
\usepackage{graphicx}
\usepackage{newtxtext}
\usepackage{newtxmath}
\usepackage{amsmath}
\usepackage{bm}
\usepackage{cite}
\usepackage[hidelinks]{hyperref}
\usepackage{eso-pic}
\usepackage{xcolor}
\usepackage{booktabs}
\usepackage{flushend}
\usepackage{resizegather}
\usepackage[font=footnotesize]{subcaption}

\graphicspath{{./figures/}}

\newcommand{\Vector}[1]{\bm{#1}}  
\newcommand{\Matrix}[1]{\bm{#1}}  
\newcommand{\Transpose}{\mathrm{T}}  
\newcommand{\refEq}[1]{(\ref{#1})}               
\newcommand{\refFig}[1]{Fig.~\ref{#1}}           
\newcommand{\refSec}[1]{Sect.~\ref{#1}}          
\newcommand{\refTable}[1]{Table~\ref{#1}}        
\newcommand{\refRem}[1]{Remark~\ref{#1}}        
\newtheorem{remark}{Remark}
\newtheorem{assumption}{Assumption}
\newcommand{\Revised}[1]{\textcolor{black}{#1}}

\hypersetup{%
    pdfauthor={Rafal Madonski, Gernot Herbst, Momir Stankovic},
    pdftitle={ADRC in Output and Error Form: Connection, Equivalence, Performance},
    pdfcreator={},pdfproducer={}
}

\begin{document}

\AddToShipoutPicture*{%
  \AtPageUpperLeft{%
    \setlength\unitlength{1cm}%
    \put(0,-0.5){\begin{minipage}[c]{\paperwidth}
    \footnotesize\centering\textcolor{black!50}{%
    This is a preprint of an article published in \emph{Control Theory and Technology}.
    The final authenticated version is available online at:} \ \textcolor{blue!60}{\url{https://doi.org/10.1007/s11768-023-00129-y}}%
    \end{minipage}}%
  }
}

\title{ADRC in Output and Error Form:\\Connection, Equivalence, Performance}

\author{Rafal Madonski\ \href{https://orcid.org/0000-0002-1798-0717}%
    {\raisebox{-0.3pt}{\includegraphics[height=9pt]{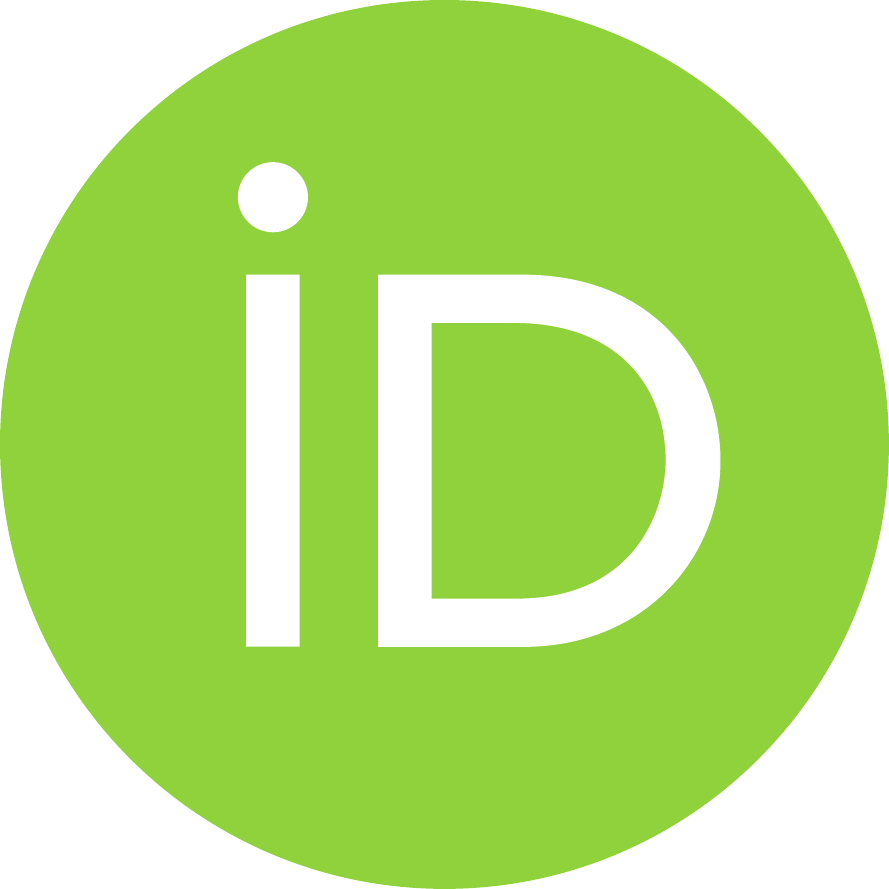}}}%
  \thanks{%
    Rafal Madonski is with the Energy and Electricity Research Center, Jinan University, China (e-mail: \protect\href{mailto:rafal.madonski@jnu.edu.cn}{rafal.madonski@jnu.edu.cn}).%
  }
  Gernot Herbst\ \href{https://orcid.org/0000-0002-4638-5378}%
    {\raisebox{-0.3pt}{\includegraphics[height=9pt]{orcid}}}%
  \thanks{%
    Gernot Herbst is with University of Applied Sciences Zwickau, Germany (e-mail: \protect\href{mailto:gernot.herbst@fh-zwickau.de}{gernot.herbst@fh-zwickau.de}).%
  }
  and
  Momir Stankovic\ \href{https://orcid.org/0000-0001-8371-9341}%
    {\raisebox{-0.3pt}{\includegraphics[height=9pt]{orcid}}}%
  \thanks{%
    Momir Stankovic is with the Military Academy, University of Defence, Belgrade, Serbia (e-mail: \protect\href{mailto:momir_stankovic@yahoo.com}{momir\_stankovic@yahoo.com}).%
  }  
}

\markboth%
{Rafal Madonski, Gernot Herbst, and Momir Stankovic: ADRC in Output and Error Form---Connection, Equivalence, Performance}
{Rafal Madonski, Gernot Herbst, and Momir Stankovic: ADRC in Output and Error Form---Connection, Equivalence, Performance}

\maketitle

\begin{abstract}
    In this work, we investigate two specific linear ADRC structures, namely output- and error-based. The former is considered a 'standard' version of ADRC, a title obtained primarily thanks to its simplicity and effectiveness, which have spurred its adoption across multiple industries. The latter is found to be especially appealing to practitioners as its feedback error-driven structure bares similarities to conventional control solutions, like PI and PID. In this paper, we describe newly found connections between the two considered ADRC structures, which allowed us to formally establish conditions for their equivalence. Furthermore, the conducted comprehensive performance comparison between output- and error-based ADRCs has facilitated the identification of specific modules within them, which can now be conveniently used as building blocks, thus aiding the control designers in customizing ADRC-based solutions and making them most suitable for their applications.
\end{abstract}


\section{Introduction}

\subsection{Background and Related Work}

Active disturbance rejection control (ADRC) is a highly practical general purpose control methodology~\cite{HanTIE}, which can be applied in a variety of ways, depending on control requirements, availability of certain signals, computational capabilities of the target hardware, etc. Especially its streamlined simplified, linear version~\cite{GaoScaling} has found its way into numerous application domains~\cite{RecentWorks}. There has been an intensified effort recently to foster the adoption of ADRC in industry and to ensure its backward compatibility with classical solutions (e.g. \cite{PID2ADRCtransit,HerbstPractical,ADRC-ZN-tuning,SaifUADRC}). This manifests itself, for example, with finding equivalence between ADRC and various well-established control schemes, like PID (\cite{ADRCpidTuning,EquivalenceSira}). Through drawing such connections, the goal is to support control practitioners by facilitating the process of ADRC understanding, designing, implementing, and tuning.

In that spirit, an error-based version of ADRC was independently proposed in~\cite{Michalekerror} and~\cite{HanPhD}, and was later generalized and proved in~\cite{GeneralError}. It allowed to rewrite the entire ADRC scheme, usually expressed in an output-based form, in a feedback error-driven form, which then enabled an immediate comparability with existing classical control solutions.\footnote{\Revised{The details of the used notation 'output-based' and 'error-based' in the context of ADRC are explained at the beginning of~\refSec{sec:ADRCschemes}.}} It also introduced a practical solution for scenarios where consecutive time-derivatives of the reference signal (needed to synthesize the feedback controller) are not available, hence cannot be used in the control law. In error-based ADRC (eADRC), they can be lumped into the total disturbance term and conveniently estimated by the extended state observer, without resorting to an additional tracking differentiator (e.g. \cite{TrackingDifferent,HanTIE}). The efficacy of eADRC was validated in numerous applications so far, including mechatronic manipulators \cite{eADRCtorsio,ESOarchitecturesKKA,ErrorMemory,ASMEgeneral}, aerial robotics \cite{eADRCuav}, power electronics \cite{CEPconverter,IJCeRESO}, process control \cite{2DOFerrorISAT}, and military applications \cite{eADRClaser}. The error-based formulation is also how the recently introduced ``ADRC Toolbox''\footnote{\url{https://www.mathworks.com/matlabcentral/fileexchange/102249-active-disturbance-rejection-control-adrc-toolbox}} for MATLAB/Simulink is implemented with~\cite{ADRCtoolbox}.

\subsection{Motivation and Goal}

One can find works, including \cite{OutputErrorComparison,SimplifyingKKA,GeneralError}, that show the differences between the oADRC and eADRC schemes in terms of control performance. Those works, however, only showed basic quantitative comparisons and did not go beyond continuous-time domain. Therefore, the current literature lacks information about the exact sources of these discrepancies and how they influence the ADRC design process as well as the realization of the control objective.

Motivated to bridge the above gap, a systematic investigation is carried in this paper with the goal of answering the question about the formal relation between oADRC and eADRC. The investigation is performed utilizing the following research methodology. First, the two control schemes are recalled in continuous-time domain for the same class of controlled systems and utilizing the same tuning methodology. Then, they are expressed as transfer functions and their respective structures are comprehensively compared using frequency domain. Such analysis allows to formally define similarities and differences between the two control schemes and better understand what is the influence of those differentiating elements on the control performance. 

Additionally, inspired by~\cite{HerbstTF}, the investigation also covers the practically important problem of realizability of the considered ADRC schemes in transfer function forms. Making sure the oADRC and eADRC schemes are realizable allows not only to conveniently analyze them in frequency domain (for example see~\cite{MomirTFadrcFPGA}), but also to compare them to existing classical solutions more easily, for example by analyzing the pole/zero placement of the feedback controller part. In return, this facilitates understanding and implementing various ADRCs using classical feedback controller and filter structures.

\subsection{Novelty}

The results of the paper contribute to the current body of knowledge related to ADRC design. The added new information can be considered as guidelines for control designers to choose a suitable ADRC-based structure and to better customize it in order to meet certain requirements in terms of implementation simplicity and control performance. The specific contributions of the paper are: 
\begin{itemize}
    \item [i)] Establishment of equivalence condition between oADRC and eADRC. A formal condition is given in frequency domain that shows the connection between the two schemes. The formed condition also shows how to go from one structure to the other, and back.
    \item [ii)] Quantitative comparison between oADRC and eADRC in time and frequency domains. Its results show what are the advantages and disadvantages of using both ADRC structures, with the special emphasis on the influence of the differentiating elements on the control performance. The systematic comparison includes key control performance metrics, like reference tracking accuracy, stability margins (gain and phase margins), disturbance rejection, and measurement noise sensitivity.
    \item [iii)] Establishment of conditions of realizability of ADRC transfer function implementation in output- and error-based forms. This allows to compare the two considered ADRCs to existing classical solutions more easily. This can be considered as an extension of~\cite{HerbstTF}, where only the output-based version was considered.
    \item [iv)] Qualitative comparison between oADRC and eADRC. Its outcome gives insight into the applicability of both structures, which is also visualized with a decision tree, dedicated to the prospective ADRC user looking for an appropriate control structure based on the available signals and desired performance. To further aid the control designers, three modules are identified that allow straightforward customization of a desired ADRC scheme.
\end{itemize}

\subsection*{Nomenclature}

Throughout the paper, we omit writing explicit dependencies from time for higher-order terms, meaning we only write $x^{(k)}$ instead of $x^{(k)}(t)$ to represent $k$-th time-derivative of signal $x(t)$. We also omit writing explicit time-dependency for dot notation for differentiation, meaning we write $\dot{x}$, $\ddot{x}$, etc., instead of redundant $\dot{x}(t)$, $\ddot{x}(t)$, etc. Additionally, symbol $\stackrel{!}{=}$ is the stacked operator indicating that both sides of the mathematical formula shall be made equal.

\section{Considered ADRC Schemes}
\label{sec:ADRCschemes}

In this section, the two considered ADRC structures, output-based and error-based, are briefly recalled. In order to allow a systematic and straightforward comparison later, they are both derived here for the same class of controlled systems. Hence, let us first introduce a $n$-th order single-input single-output plant model with output $y(t)$, control input $u(t)$, and $d(t)$ being the external disturbance input:
\begin{equation}
    y^{(n)}=\sum_{i=0}^{n-1}a_i \cdot y^{(i)}+b \cdot u(t)+d(t),
 \label{eqn:General_nth_order_plant}   
\end{equation}
where $a_i$ are generally unknown plant parameters and $b$ is a partially known plant gain. Since ADRC is a general purpose methodology, it can be designed in various ways and tailor made to a particular class of systems and specific application. That is why, in this section, we specify which exact ADRC variants we are considering in the study. Therefore, the following clarifying terminology is introduced. 

\textbf{Output-based ADRC (oADRC)}. This is the commonly seen version of ADRC in the literature, popularized by Prof. Zhiqiang Gao through his seminal work~\cite{GaoScaling}. It is often considered the \textit{classic} version of ADRC. It utilizes a linear observer (extended with one extra state variable, representing the total disturbance) and a linear state feedback controller. It is denoted in this paper as "output-based" since the system's state vector in this approach, used during the observer and controller syntheses, consists of plant's output $y(t)$ and its derivatives $\dot{y},\ddot{y},\ldots,y^{(n-1)}$. A block diagram of oADRC can be seen in~\refFig{fig:ADRC_time_domain}.

\textbf{Error-based ADRC (eADRC)}. This is a version of the ADRC scheme that, similarly to oADRC utilizes a linear observer (extended with one extra state variable, representing the total disturbance) and a linear state feedback controller, but the system's state vector in this instance, used during the observer and controller syntheses, consists of feedback error $e(t)\triangleq r(t)-y(t)$ and its consecutive time derivatives $\dot{e},\ddot{e},\ldots,e^{(n-1)}$. It can be viewed as a special form of the classic oADRC. More detailed information on the reasoning behind such representation and some exemplary applications can be found in~\cite{Michalekerror,GeneralError}. A block diagram of eADRC can be seen in~\refFig{fig:eADRC_time_domain}.

\begin{remark}
    The above terminology oADRC and eADRC is not widely used in the scientific literature but is used here specifically and locally to distinguish the two types of ADRC structures considered in this paper.
\end{remark}

Next, the considered ADRC schemes are derived, oADRC in \refSec{sec:OutputBasedADRC} and eADRC in \refSec{sec:ErrorBasedADRC}, which lays the foundation of their comparative analysis in \refSec{sec:CompareAnalysis}. 

\subsection{Output-based ADRC (oADRC)}
\label{sec:OutputBasedADRC}

By including uncertainties of the plant gain as $b=b_0+\Delta{b}$, where $b_0 \neq 0$ is the best available estimate of the $b$ and $\Delta{b}$ is the parametric uncertainty, the application of the ADRC methodology allows to represent~\refEq{eqn:General_nth_order_plant} in a more compact form:
\begin{equation}
    y^{(n)}=f(t)+b_0 \cdot u(t),
 \label{eqn:ADRC_representation_of_general_nth_order_plant}
\end{equation}
where $f(t)=\sum_{i=0}^{n-1}a_i \cdot y^{(i)}+\Delta{b} \cdot u(t)+d(t)$ is denoted as the generalized (total) disturbance.

A block diagram of the continuous-time linear ADRC in its original state-space output-based form for the plant model~\refEq{eqn:ADRC_representation_of_general_nth_order_plant} is presented in~\refFig{fig:ADRC_time_domain}. Its main components include: plant gain inversion, disturbance rejection utilizing the estimated total disturbance term, and an outer state-feedback controller. The control law is formed as:
\begin{figure}[t]
    \centering
    \includegraphics[width=\columnwidth]{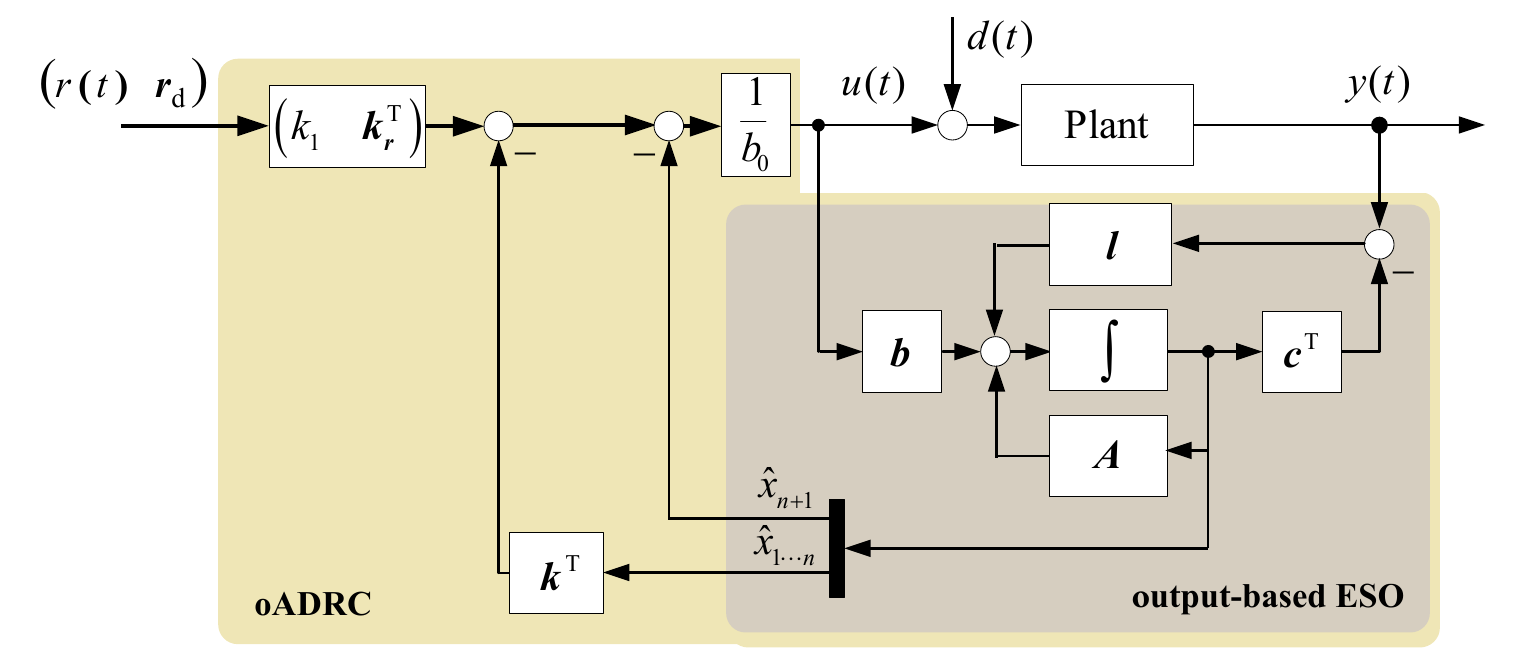}
    \caption{Continuous-time output-based ADRC (oADRC) in state-space implementation. \Revised{Here we denote the observer as output-based ESO as one of its inputs is the system output signal $y(t)$. The second input is the control signal $u(t)$, which is also the case for observer in \refFig{fig:eADRC_time_domain}.}}
     \label{fig:ADRC_time_domain}
 \end{figure}
\begin{equation}
    u(t) = \frac{1}{b_0} \cdot \left( k_1 \cdot r(t)+ \Vector{k_\textrm{r}}^\Transpose \cdot \Vector{r_\textrm{d}}- \begin{pmatrix}
        \Vector{k}^\Transpose  &  1
    \end{pmatrix} \cdot \Vector{\hat{x}} \right),
    \label{eqn:ADRC_CT_SS_Controller}
\end{equation}
with \quad $\Vector{k_\textrm{r}}^\Transpose =
    \begin{pmatrix}
        k_2  &  \cdots  &  k_n & 1
    \end{pmatrix}$,
    $\Vector{k}^\Transpose =
    \begin{pmatrix}
       k_1   & \cdots & k_n
    \end{pmatrix}$,\\ 
    and \quad $\Vector{r_\textrm{d}}^\Transpose =
    \begin{pmatrix}
        \dot{r} &  \cdots   &  r^{(n)}
    \end{pmatrix}$.
    
The state estimation vector, defined as:
\begin{align}
    \Vector{\hat{x}} &= \begin{pmatrix}
        \hat{x}_1(t)  &  \cdots & \hat{x}_n(t) &  \hat{x}_{n+1}(t)
    \end{pmatrix}^\Transpose\notag \\
    &=\begin{pmatrix}
        \hat{y}(t)  &  \cdots  & \hat{y}^{(n-1)} & \hat{f}(t)
    \end{pmatrix}^\Transpose,\notag
\end{align}
can be obtained using the extended state observer (ESO):
\begin{equation}
    \Vector{\dot{\hat{x}}}
    = \Matrix{A} \cdot \Vector{\hat{x}}
    + \Vector{b} \cdot u(t)
    + \Vector{l} \cdot \left( y(t) - \Vector{c}^\Transpose \cdot  \Vector{\hat{x}} \right),
    \label{eqn:ADRC_CT_SS_Observer}
\end{equation}
\vspace*{-.5cm}
\begin{align}
    \text{with} \quad \Matrix{A} &= \begin{pmatrix}
        \Matrix{0}^{n \times 1}  &  \Matrix{I}^{n \times n}  \\
        0  &  \Matrix{0}^{1 \times n}
    \end{pmatrix}
    ,\quad
    \Vector{b} = \begin{pmatrix}
        \Matrix{0}^{(n-1) \times 1}  \\
        b_0  \\
        0
    \end{pmatrix},\notag\\
    \Vector{l} &= \begin{pmatrix}
        l_1  &  \cdots  &  l_{n+1}
    \end{pmatrix}^\Transpose
    ,\quad
    \Vector{c}^\Transpose = \begin{pmatrix}
        1  &  \Matrix{0}^{1 \times n}
    \end{pmatrix}.\notag
\end{align}

The application of oADRC scheme leads to a robust feedback-linearization of the original system dynamics~\refEq{eqn:General_nth_order_plant}, which is followed next by a design of the outer-loop state-feedback with feedforward controller designed for the resultant chain of integrators (see \refFig{fig:ADRC_time_domain}). One can notice that the compensation of total disturbance $f(t)$ can be seen as an inner control loop with the actual plant. The inner control loop forms a canonical plant for the outer state-feedback controller $\Vector{k}^\Transpose$ and has a unity-gain integrator chain behavior.

\begin{remark}
    The use of feedforward term in the control law~\refEq{eqn:ADRC_CT_SS_Controller} allows to give the desired steady-state value. If the reference signals are available, the generated feedforward signal is useful in shaping the system response~\cite{ACCerror}.
\end{remark}

\begin{remark}
\label{lab:CaseRefDerivatives}
    It should be noted that the control law~\refEq{eqn:ADRC_CT_SS_Controller} can also be realized without the reference derivatives (i.e. $\Vector{r_\textrm{d}}=0$) as shown in \cite{GaoScaling,GernotElectronics}). For the purpose of this work, such scenario, where the reference derivatives are not available for controller synthesis, will be considered as the nominal version of ADRC and denoted as 'case A'. Alternatively, a scenario, where all the reference derivatives are available, will be denoted as 'case B'. It is a deliberate choice of ours to focus on cases A and B as they represent opposite \textit{extremes}, where either none or all of the reference derivatives are available for controller synthesis. It is intuitive that if the information about the reference derivative is somehow available, either through a'priori knowledge or on-line calculation using a signal differentiator (e.g. \cite{HanTIE}), then such information should be used as it can improve control quality. To visualize this relation, in the~Appendix, we have prepared a generic example of a third-order system ($n=3$) to show the influence of different levels of knowledge about the reference derivatives on the tracking quality.
\end{remark}     

\paragraph{Controller Tuning} Assuming the observer's convergence is fast enough, the rejection of total disturbance will be so effective that the (outer) state-feedback controller can control canonical plant model (denoted using subscript 'CP') with unity-gain integrator chain behaviour, regardless of the actual plant. By defining a system state vector as $\Vector{x}_\mathrm{CP}=\begin{pmatrix}y(t) & \cdots& y^{(n-1)}\end{pmatrix}$, the state-space model of the canonical plant with unity-gain integrator chain can be expressed as:
\begin{gather}
\Vector{\dot{x}}_\mathrm{CP}
    = \Matrix{A}_\mathrm{CP} \cdot \Vector{x}_\mathrm{CP}
    + \Vector{b}_\mathrm{CP} \cdot u(t),    
    \\
    \text{with}\quad
    \Matrix{A}_\mathrm{CP} = \begin{pmatrix}
        \Matrix{0}^{(n-1) \times 1}  &  \Matrix{I}^{(n-1) \times (n-1)}  \\
        0  &  \Matrix{0}^{1 \times (n-1)}
    \end{pmatrix}
    ,\quad
    \Vector{b}_\mathrm{CP} = \begin{pmatrix}
        \Matrix{0}^{(n-1) \times 1}  \\
        1
    \end{pmatrix}.
    \label{ADRC_virtual_plant_control}\nonumber
\end{gather}
Now, by applying the `bandwidth parameterization' approach from~\cite{GaoScaling}, which consists of placing all the poles at one common location $\lambda = -\omega_\mathrm{CL}$, makes the controller tuning relatively simple as it comes down to choosing only the desired closed-loop bandwidth $\omega_\mathrm{CL}$, i.e.:
\begin{align}
    \left( \lambda + \omega_\mathrm{CL} \right)^n
    &\stackrel{!}{=}
    \det\left( \lambda \Matrix{I} - \left( \Matrix{A}_\mathrm{CP} - \Vector{b}_\mathrm{CP} \Vector{k}^\Transpose \right) \right)\notag\\
    &= \lambda^{n} + k_n \lambda^{n-1} + \ldots + k_2 \lambda + k_1
    .
    \label{eqn:ADRC_CT_SS_Controller_K_PolePlacement_OutputBased}
\end{align}
The binominal expansion of \refEq{eqn:ADRC_CT_SS_Controller_K_PolePlacement_OutputBased} leads to the controller gains:
\begin{equation}
    k_i = \frac{n!}{(n-i+1)! \cdot (i-1)!} \cdot \omega_\mathrm{CL}^{n-i+1},
    \quad \forall i = 1, \ldots, n.
    \label{eqn:ADRC_CT_SS_Controller_K_OutputBased}
\end{equation}

\begin{remark}
\label{rem:BWtuning}
    The use of bandwidth parameterization for tuning is a subjective choice of ours, driven mostly by its simplicity as well as its prevalence in the scientific literature on ADRC. A variety of other tuning methods could be applied here as well, with handful examples shown in~\cite{surveyISAT}.
\end{remark}

\paragraph{Observer Tuning} In the same manner as in the controller tuning, the observer gains can be obtained, for example, based on the bandwidth parameterization approach (\cite{GaoScaling}) by placing the closed-loop observer poles at one common location $\lambda = -k_\mathrm{ESO} \cdot \omega_\mathrm{CL}$, with $k_\mathrm{ESO}$ being the relative factor between the observer loop and the control loop bandwidths:
\begin{align}
    \left( \lambda + k_\mathrm{ESO} \cdot \omega_\mathrm{CL} \right)^{n+1}
    &\stackrel{!}{=}
    \det\left( \lambda \Matrix{I} - \left( \Matrix{A} - \Vector{l} \Vector{c}^\Transpose \right) \right)\notag\\
    &= \lambda^{n+1} + l_1 \lambda^{n} + \ldots + l_{n} \lambda + l_{n+1}
    .
    \label{eqn:ADRC_CT_SS_Observer_L_PolePlacement_OutputBased}
\end{align}
As in the case of computing the controller gains, the binominal expansion of \refEq{eqn:ADRC_CT_SS_Observer_L_PolePlacement_OutputBased} yields the observer gains:
\begin{equation}
    l_i = \frac{(n+1)!}{(n-i+1)! \cdot i!} \cdot \left( k_\mathrm{ESO} \cdot \omega_\mathrm{CL} \right)^i,
    \quad \forall i = 1, \ldots, n+1
    .
    \label{eqn:ADRC_CT_SS_Observer_L_OutputBased}
\end{equation}

\subsection{Error-based ADRC (eADRC)}
\label{sec:ErrorBasedADRC}

In \refSec{sec:OutputBasedADRC}, it was shown that in case~B of oADRC, deriving the outer-loop state-feedback with feedforward controller designed for the resultant chain of integrators assumes perfect knowledge (availability) of the consecutive reference time-derivatives up to the order of the linearized dynamics. Through the use of eADRC, this assumption can be removed, as it allows to treat simultaneously the reference trajectory as a source of an additional external disturbance which has to be actively rejected~\cite{Michalekerror,HanPhD,GeneralError}.

\begin{figure}[t]
    \centering
    \includegraphics[width=\columnwidth]{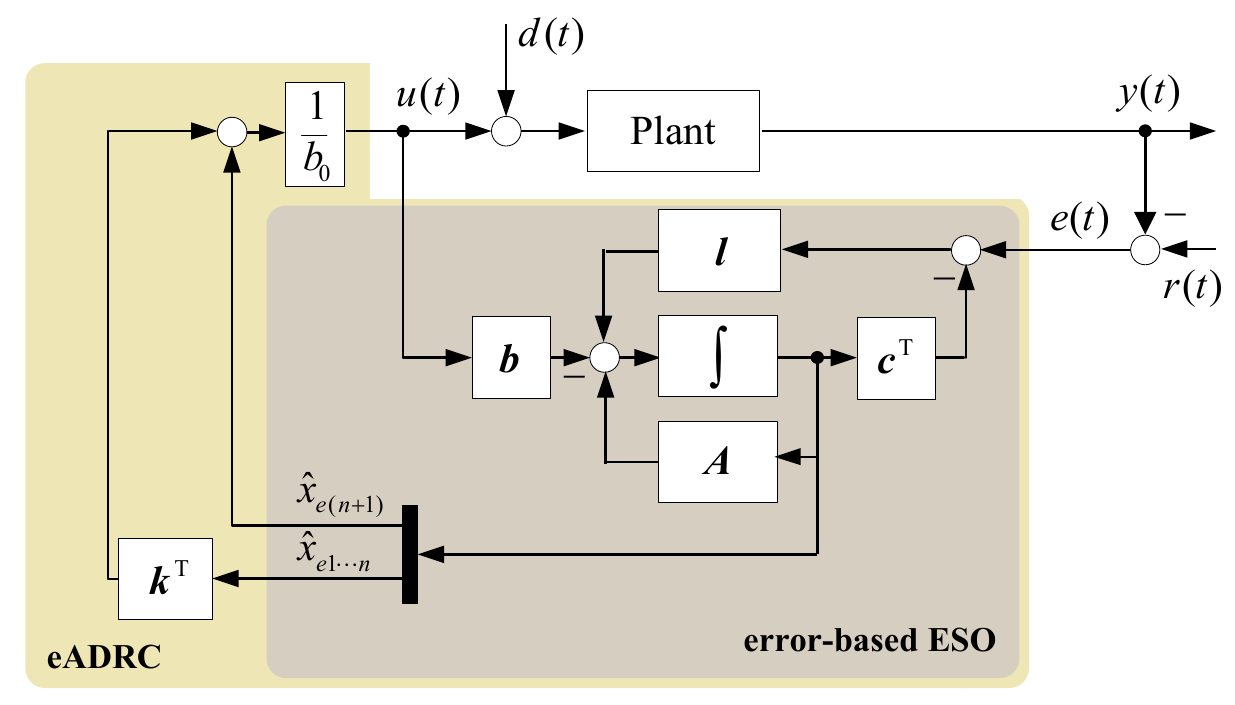}
    \caption{Continuous-time error-based ADRC (eADRC) in state-space implementation. \Revised{Here we denote the observer as error-based ESO as one of its inputs is the feedback error signal $e(t)$. The second input is the control signal $u(t)$, which is also the case for observer in \refFig{fig:ADRC_time_domain}.}}
     \label{fig:eADRC_time_domain}
 \end{figure}

Following the definition of the tracking error $e(t) \triangleq r(t) - y(t)$, system~\refEq{eqn:General_nth_order_plant} can be reformulated as:
\begin{equation}
    e^{(n)}=f_{\textrm{e}}(t)-b_0 \cdot u(t),
 \label{eqn:eADRC_representation_of_general_nth_order_plant}   
\end{equation}
where $f_{\textrm{e}}(t)=r^{(n)}-\sum_{i=0}^{n-1}a_i \cdot r^{(i)}+\sum_{i=0}^{n-1}a_i \cdot e^{(i)}- \Delta{b} \cdot u(t)-d(t)$ is the generalized (total) disturbance in the error domain (cf.~\refEq{eqn:ADRC_representation_of_general_nth_order_plant}).

\begin{remark}
     In \refEq{eqn:eADRC_representation_of_general_nth_order_plant}, the initial reference trajectory tracking control task is reformulated into a regulation control task, hence the reference signal and its derivatives can become part of the total disturbance~$f_{\textrm{e}}(t)$.
\end{remark}

A block diagram of the eADRC structure is shown in \refFig{fig:eADRC_time_domain}. One can see that the main difference, compared to the standard output-based form depicted in~\refFig{fig:ADRC_time_domain}, is the feedback controller, here based on the tracking error estimation vector:
\begin{equation}
    u_\textrm{e}(t) = \frac{1}{b_0} \cdot \left(  \begin{pmatrix}
        \Vector{k}^\Transpose  &  1
    \end{pmatrix} \cdot \Vector{\hat{x}_\textrm{e}} \right),
    \label{eqn:eADRC_CT_SS_Controller}
\end{equation}
\vspace*{-.5cm}
\begin{align}
    \text{where} \quad \Vector{\hat{x}_\textrm{e}} &= \begin{pmatrix}
        \hat{x}_{\textrm{e}1}(t)  &  \cdots  & \hat{x}_{\textrm{e}(n)}(t) & \hat{x}_{\textrm{e}(n+1)}(t)
    \end{pmatrix}^\Transpose\notag\\
    &= \begin{pmatrix}
        \hat{e}(t)  &  \cdots  & \hat{e}^{(n-1)} & \hat{f}_\textrm{e}(t)
    \end{pmatrix}^\Transpose,\notag
\end{align}
is the state estimation vector, obtained by the following error-based version of the ESO:
\begin{equation}
    \Vector{\dot{\hat{x}}_\textrm{e}}
    = \Matrix{A} \cdot \Vector{\hat{x}_\textrm{e}}
    - \Vector{b} \cdot u_\textrm{e}(t)
    + \Vector{l} \cdot \left( e(t) - \Vector{c}^\Transpose \cdot  \Vector{\hat{x}_\textrm{e}} \right).
    \label{eqn:eADRC_CT_SS_Observer}
\end{equation}

It can be seen that the above error-based ESO has almost the same structure as the output-based ESO in \refEq{eqn:ADRC_CT_SS_Observer}, with one of the differences being the sign of the control signal-related term $\Vector{b} \cdot u_\textrm{e}(t)$. Also, in~\refEq{eqn:ADRC_CT_SS_Observer}, the input is the system output $y(t)$, here, the input is the tracking error $e(t)$.


\paragraph{Controller and Observer Tuning} 
Assuming perfect rejection of the total disturbance $f_{\textrm{e}}(t)$, in the same manner as in standard ADRC in \refSec{sec:OutputBasedADRC}, the controller design and tuning can be performed for a canonical plant with unity-gain integrator chain. Then, having in mind \refRem{rem:BWtuning}, the same `bandwidth parameterization' approach from~\cite{GaoScaling} can be applied in eADRC as well. Hence, choosing the desired closed-loop bandwidth $\omega_\mathrm{CL}$ leads to the same controllers coefficients as in \refEq{eqn:ADRC_CT_SS_Controller_K_PolePlacement_OutputBased}. Similarly, following the procedure from~\refEq{eqn:ADRC_CT_SS_Observer_L_PolePlacement_OutputBased}, the observer gains can be calculated as in  \refEq{eqn:ADRC_CT_SS_Observer_L_OutputBased}. This means that the eADRC uses the same tuning methodology as the output-based structure, which retains simplicity of only two tuning parameters ($\omega_\mathrm{CL}$ and $k_\mathrm{ESO}$), which is one of the selling points of ADRC to industry.

\section{Comparative Analysis}
\label{sec:CompareAnalysis}

In this section, a comparison is performed between oADRC (\refSec{sec:OutputBasedADRC}) and eADRC (\refSec{sec:ErrorBasedADRC}) in order to answer three fundamental questions about their relation. The first one, in \refSec{sec:Q1relation}, is about establishing a formal relation between the two approaches, with formal definitions of their similarities and differences. The second question, in \refSec{sec:Q2influence}, is about the influence of their differentiating elements on the control performance. The third one, in \refSec{sec:Q3realizability}, is about their realizability in practice. For the purpose of the comparison, following assumptions are made.
\begin{assumption}
    In the comparison, we consider the same class of systems defined as seen in~\refEq{eqn:General_nth_order_plant}. Special focus is on first- and second-order systems due to their prevalence in engineering practice. 
\end{assumption}
\begin{assumption}
    To make the comparison fair, the same observer and controller tuning methodology is used, namely the single bandwidth parameterization approach from~\cite{GaoScaling}.
\end{assumption}
\begin{assumption}
    For oADRC, two practical cases are considered (see~\refRem{lab:CaseRefDerivatives}), one assuming the availability of the reference signal derivatives for controller synthesis (case~B) and another without it (case~A). For eADRC, no extra cases need to be defined since this scheme does not require the reference signal derivatives for the design process. The control schemes and their specific cases used in the upcoming analysis are listed in~\refTable{tab:ConsideredControlSchemes}.
\end{assumption}
\begin{table}[htb!]
\centering
\caption{The investigated control schemes.}
\label{tab:ConsideredControlSchemes}
{\renewcommand{\arraystretch}{1.5}
\begin{tabular}{ccc}
\hline
\textbf{Control scheme} & \textbf{Gains}  & \textbf{Assumption}                                                         \\ \hline
oADRC (case A) & $\Vector{k_\textrm{r}}=0$ & $\Vector{r_\textrm{d}}$ not available     \\ 
oADRC (case B) &   $\Vector{k_\textrm{r}}\neq 0$   & $\Vector{r_\textrm{d}}$ available \\ 
eADRC & \textbf{---}                    & $\Vector{r_\textrm{d}}$ not required    \\ \hline
\end{tabular}%
}
\end{table}

\subsection{What is the Relation Between oADRC and eADRC?}
\label{sec:Q1relation}

Based on the preliminary information about the two considered ADRC structures in Sect.~\ref{sec:ADRCschemes}, one can notice that eADRC scheme does not explicitly use the reference channel in this derivation but still bares significant similarities with its oADRC counterpart. By visual examination of \refFig{fig:ADRC_time_domain} and \refFig{fig:eADRC_time_domain}, following observations can be made. From the oADRC perspective, the eADRC can be viewed as its inner component. On the other hand, eADRC can be viewed as a simplified, \textit{bare-bones} version of oADRC---its special case when the reference channel is not explicitly used for controller synthesis. Therefore, there seems to be a connection between the two ADRC schemes.

To enable better understanding of the considered oADRC and eADRC structures, and potentially find formal conditions of their equivalence, we first introduce transfer-function representation of their continuous-time forms. We start with the oADRC. The Laplace transform of the control signal \refEq{eqn:ADRC_CT_SS_Controller} is substituted in \refEq{eqn:ADRC_CT_SS_Observer}, which gives\footnote{Throughout the paper, we assume, without loss of generality, that the initial values that appear when making Laplace transformation are equal to zero.}:
%
%
\begin{equation}
    \Vector{\hat{x}}(s)
    = \left(s \Matrix{I} - \Matrix{A}_\mathrm{CL} \right)^{-1}
    \cdot \left[ \frac{\Vector{b}}{b_0} \left(k_1\cdot r(s) + \Vector{k_\textrm{r}}^\Transpose \cdot \Vector{r_\textrm{d}}(s)\right) + \Vector{l} \cdot y(s) \right]
    \label{eqn:ADRC_CT_TF_Transfer_X}
\end{equation}
where:
\begin{align}
    \quad \Matrix{A}_\mathrm{CL}
    &=
    \Matrix{A} - \Vector{l}\Vector{c}^\Transpose - \frac{1}{b_0} \Vector{b} \begin{pmatrix}
     \Vector{k}^\Transpose   & 1 
    \end{pmatrix}\label{eqn:ADRC_CT_TF_Transfer_ACL}\\
    &=
    \begin{pmatrix}
        \Matrix{0}^{n \times 1}  &  \Matrix{I}^{n \times n}
        \\
        0  &  \Matrix{0}^{1 \times n}
    \end{pmatrix}
    -
    \begin{pmatrix}
        \Vector{l}  &  \Matrix{0}^{(n+1) \times n}
    \end{pmatrix}
    -
    \begin{pmatrix}
        \Matrix{0}^{(n-1) \times (n+1)}
        \\
        \begin{pmatrix}
            \Vector{k}^\Transpose  &  1 
        \end{pmatrix}
        \\
        \Matrix{0}^{1 \times (n+1)}
    \end{pmatrix}.\notag
\end{align}
We eliminate the observer estimation vector $\hat{\Vector{x}}$ by substituting \refEq{eqn:ADRC_CT_TF_Transfer_X} in the Laplace transform of \refEq{eqn:ADRC_CT_SS_Controller}, which yields:
\begin{align}
    u(s) &= \frac{k_1}{b_0} r(s)+\frac{1}{b_0}\cdot \Vector{k_\textrm{r}}^\Transpose\cdot \Vector{r_\textrm{d}}(s) - \frac{1}{b_0} \begin{pmatrix}
        \Vector{k}^\Transpose & 1
    \end{pmatrix} \cdot \label{eqn:ADRC_CT_TF_Transfer_u(s)}\\
    &\cdot\left(s \Matrix{I} - \Matrix{A}_\mathrm{CL}\right)^{-1}
    \cdot\left[ \frac{\Vector{b}}{b_0}\left( k_1 r(s) + \Vector{k_\textrm{r}}^\Transpose \Vector{r_\textrm{d}}(s)\right)+  \Vector{l} \cdot y(s) \right].\notag
\end{align}

By introducing $\Vector{r_\textrm{d}}(s)=r(s)\cdot \begin{pmatrix}
    s & \cdots & s^{(n)}
\end{pmatrix}^\Transpose$, the control signal \refEq{eqn:ADRC_CT_TF_Transfer_u(s)} can be rewritten as seen in~\refEq{eqn:ADRC_CT_TF_Transfer_form}, 
\begin{table*}[htb!]
\centering
\begin{minipage}{\textwidth}
    \normalsize \begin{equation}
        u(s)
        =\frac{1}{b_0} \begin{pmatrix}
            \Vector{k}^\Transpose & 1
        \end{pmatrix} \cdot \left(s \Matrix{I} - \Matrix{A}_\mathrm{CL}\right)^{-1} \Vector{l}\cdot  \left[\left(\frac{\left(b_0-  \begin{pmatrix}
            \Vector{k}^\Transpose & 1
        \end{pmatrix} \cdot \left(s \Matrix{I} - \Matrix{A}_\mathrm{CL}\right)^{-1}\Vector{b}\right) \cdot \left(k_1+\Vector{k_\textrm{r}}^\Transpose \begin{pmatrix}
            s &\cdots & s^{(n)}
        \end{pmatrix}^\Transpose\right)}
        {b_0\begin{pmatrix}
            \Vector{k}^\Transpose & 1
        \end{pmatrix} \cdot \left(s \Matrix{I} - \Matrix{A}_\mathrm{CL}\right)^{-1}\Vector{l}}\right)r(s)-y(s)\right] 
        \label{eqn:ADRC_CT_TF_Transfer_form}
    \end{equation}
\medskip
\hrule
\end{minipage}
\end{table*}
which is equivalent to the standard transfer function\footnote{Due to the size of the equation, it has been written separately on next page while keeping chronological numbering.}:
\begin{equation}
    u(s) = G_\mathrm{FB}(s) \cdot \left[ G_\mathrm{PF}(s) \cdot r(s) - y(s) \right], 
    \label{eqn:ADRC_CT_TF_Representation}
 \end{equation}
where:
\begin{equation}
    G_{\mathrm{FB}}(s) =\frac{1}{b_0} \cdot \begin{pmatrix}
        \Vector{k}^\Transpose  &  1
    \end{pmatrix} \frac{\operatorname{adj}\left( s \Matrix{I} - \Matrix{A}_\mathrm{CL} \right)}{\operatorname{det}\left( s \Matrix{I} - \Matrix{A}_\mathrm{CL} \right)} \cdot \Vector{l},
    \label{eqn:Feedback_TF}
    \end{equation}
is the feedback transfer function while \refEq{eqn:PreFilter_TF_with_ref_derivatives}
%
\begin{table*}[htb!]
\centering
\begin{minipage}{\textwidth}
    \normalsize     \begin{equation}   
        G_{\mathrm{PF}}(s) =
        \frac{\left(k_1+\Vector{k_\textrm{r}}^\Transpose\begin{pmatrix}
            s  &\cdots & s^{(n)}
        \end{pmatrix}^\Transpose\right)\left(\operatorname{det}\left( s \Matrix{I} - \Matrix{A}_\mathrm{CL} \right)-\frac{1}{b_0}\begin{pmatrix}
            k^\Transpose & 1
        \end{pmatrix} \operatorname{adj}\left( s \Matrix{I} - \Matrix{A}_\mathrm{CL} \right)\Vector{b}\right)}{\begin{pmatrix}
            k^\Transpose & 1
        \end{pmatrix} \operatorname{adj}\left( s \Matrix{I} - \Matrix{A}_\mathrm{CL} \right)\cdot \Vector{l}}
       \label{eqn:PreFilter_TF_with_ref_derivatives}
    \end{equation}
\medskip
\hrule
\end{minipage}
\end{table*}
represents the reference pre-filter transfer function, derived for the control law \refEq{eqn:ADRC_CT_SS_Controller} with the appropriate reference derivatives (case~B). The pre-filter transfer function, defined by \refEq{eqn:PreFilter_TF_with_ref_derivatives}, can be expressed alternatively as$^{\text{2}}$:
\begin{equation}   
    G_{\mathrm{PF}}(s) =G_{\mathrm{R}}(s)\cdot  \bar{G}_{\mathrm{PF}}(s),
   \label{eqn:PreFilter_TF_with_ref_derivatives_reformulation}
\end{equation}
with:
\begin{gather}   
    \bar{G}_{\mathrm{PF}}(s) =
    \frac{\operatorname{det}( s \Matrix{I} - \Matrix{A}_\mathrm{CL}) -\frac{1}{b_0}\begin{pmatrix}
        k^\Transpose & 1
    \end{pmatrix} \operatorname{adj}\left( s \Matrix{I} - \Matrix{A}_\mathrm{CL} \right)\Vector{b}}{\begin{pmatrix}
        k^\Transpose & 1
    \end{pmatrix} \operatorname{adj}\left( s \Matrix{I} - \Matrix{A}_\mathrm{CL} \right)\cdot \Vector{l}}
    ,
   \label{eqn:PreFilter_TF_without_ref_derivatives}
\end{gather}
and $G_{\mathrm{R}}(s) =k_1+\Vector{k_\textrm{r}}^\Transpose\begin{pmatrix}
        s  &\cdots & s^{(n)}
\end{pmatrix}^\Transpose$, where the latter results from summing the reference signal and its derivatives and multiplying them with appropriate gains. The pre-filter transfer function for the control law~\refEq{eqn:ADRC_CT_SS_Controller}, generated without the reference derivatives (case~A), can be simply obtained from  \refEq{eqn:PreFilter_TF_with_ref_derivatives_reformulation} by substituting $G_{\mathrm{R}}(s) =k_1$. It should be noted, however, that both cases A and B have the same feedback transfer function \refEq{eqn:Feedback_TF}, because it does not depend on the reference derivatives.

Now, to represent eADRC in the transfer function form, similar approach is taken. The Laplace transform of the control signal \refEq{eqn:eADRC_CT_SS_Controller} is first substituted in \refEq{eqn:eADRC_CT_SS_Observer}, which gives:   
\begin{equation}
    \Vector{\hat{x}_\textrm{e}}(s)
    = \left(s \Matrix{I} -  \Matrix{A}_\mathrm{CL} \right)^{-1}
    \cdot \Vector{l}\cdot e(s)
    .
    \label{eqn:eADRC_CT_TF_Transfer_X}
\end{equation}
By substituting \refEq{eqn:eADRC_CT_TF_Transfer_X} in the Laplace transform of \refEq{eqn:eADRC_CT_SS_Observer} and eliminating $\Vector{\hat{x}}_\textrm{e}(s)$, eADRC transfer function is: 
\begin{equation}
    u(s)=\frac{1}{b_0} \cdot \begin{pmatrix}
        \Vector{k}^\Transpose  &  1
    \end{pmatrix} \frac{\operatorname{adj}\left( s \Matrix{I} - \Matrix{A}_\mathrm{CL} \right)}{\operatorname{det}\left( s \Matrix{I} - \Matrix{A}_\mathrm{CL} \right)} \cdot \Vector{l}\cdot e(s).
    \label{eqn:eADRC_CT_TF_Transfer_U}
\end{equation} 
It is evident that~\refEq{eqn:eADRC_CT_TF_Transfer_U} can be represented with only the feedback transfer function from the tracking error ($e$) to the control input ($u$):
\begin{equation}
    \frac{u(s)}{e(s)}=\frac{1}{b_0} \cdot \begin{pmatrix}
        \Vector{k}^\Transpose  &  1
    \end{pmatrix} \frac{\operatorname{adj}\left( s \Matrix{I} - \Matrix{A}_\mathrm{CL} \right)}{\operatorname{det}\left( s \Matrix{I} - \Matrix{A}_\mathrm{CL} \right)} \cdot \Vector{l}\equiv G_{\textrm{FB}}(s),
    \label{eqn:eADRC_CT_TF_Transfer_Umod}
\end{equation} 
which is identical to the oADRC feedback transfer function $G_{\textrm{FB}}(s)$ from~\refEq{eqn:Feedback_TF}.

\begin{figure}[htb!]
    \centering
    \includegraphics[width=\columnwidth]{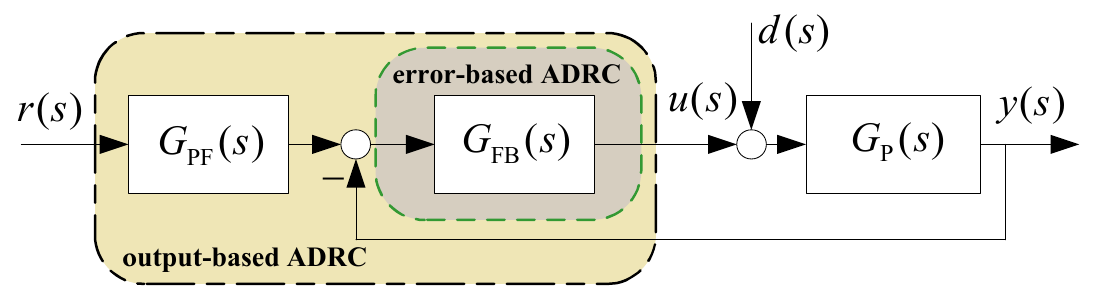}
    \caption{Transfer functions of oADRC and eADRC variants.}
     \label{fig:ADRC_eADRC_tf_forms}
 \end{figure}

This established relation between oADRC and eADRC control systems of the plant model $G_{\textrm{P}}(s)$ can be visually expressed through \refFig{fig:ADRC_eADRC_tf_forms}. From this figure, one can notice that the entire eADRC has the same structure as the feedback transfer function of oADRC, or looking at it from a different perspective, when the oADRC structure has no pre-filter (i.e. $G_{\textrm{PF}}(s) = 1$), it boils down to eADRC.

\begin{remark}
    It should be noted that oADRC, in the form shown in \refSec{sec:OutputBasedADRC}, although (visually) has a structure of a two degrees-of-freedom (2DOF) system with transfer functions $G_{\textrm{FB}}(s)$ and $G_{\textrm{PF}}(s)$, is not strictly a 2DOF control scheme. That is of course if one follows the definition from~\cite{araki2003two}, which states that a controller is of 2DOF structure if one can write it using two transfer functions that can be tuned fully independently. The used oADRC form does not satisfy that definition because its transfer functions are coupled with each other through the use of tuning parameters $k_\mathrm{ESO}$ and $\omega_\mathrm{CL}$, hence it is not possible to decouple setting of the disturbance rejection and set-point tracking performances, which is a strict requirement given the above definition of a 2DOF control system. The eADRC, on the other hand, has a typical one degree-of-freedom (1DOF) control structure.
\end{remark}

\textbf{Answer}. To address the question posed in the title of this subsection, the direct comparison of output- and error-based control algorithms, conducted in $s$-domain, shows that the pre-filter transfer function $G_\mathrm{PF}(s)$ is the element that differentiates the two designs (see \refFig{fig:ADRC_eADRC_tf_forms}). The relation between oADRC and eADRC can be thus viewed from two perspectives. On one hand, the eADRC can be considered as a special instance of oADRC when $G_\mathrm{PF}(s)=1$. On the other hand, the oADRC can be considered as an extended control structure, where additional component (namely $G_\mathrm{PF}(s)$) is added to the eADRC. It is thus interesting to investigate the influence of the term $G_\mathrm{PF}(s)$ on the system performance using oADRC with respect to eADRC. Hence, this question is investigated next.

\begin{remark}
    The equivalence between oADRC and eADRC structures, established in this subsection, is also valid in discrete time. A detailed derivation of that fact is, however, omitted here to avoid redundancy and the necessity of having to introduce another domain, related to discrete time.
\end{remark}

\subsection{What is the Influence of Term $G_\mathrm{PF}(s)$ on the System Performance?}
\label{sec:Q2influence}

It was revealed in \refSec{sec:Q1relation} that the pre-filter transfer function $G_\mathrm{PF}(s)$ is a component that fundamentally differentiates the considered oADRC and eADRC structures from each other. Therefore, the influence of $G_\mathrm{PF}(s)$ on the system stability as well as on steady-state and transient performances, is analyzed here. 

\paragraph{Stability and Steady-State Performance} Based on the previously derived transfer functions of the considered oADRC and eADRC structures (cf. \refFig{fig:ADRC_eADRC_tf_forms}), it is evident that their open-loop transfer functions are the same, i.e.:
\begin{equation}
    G_\mathrm{OL}(s)=G_\mathrm{FB}(s)G_{\textrm{P}}(s),
    \label{eqn:Open_loop_tf}
\end{equation}
which means that they have the same stability characteristics (for the same tuning parameters), regardless of $G_\mathrm{PF}(s)$.

One can also see that oADRC and eADRC structures have identical transfer functions from disturbance $d(s)$ to output $y(s)$:
\begin{equation}
    G_\mathrm{YD}(s)=\frac{G_{\textrm{P}}(s)}{1+G_\mathrm{OL}(s)}, 
    \label{eqn:dist_rejection_tf}
\end{equation}
which means that they have the same disturbance rejection capabilities in steady-state.

In both considers ADRC schemes, the transfer function between the measurement noise $n(s)$ in the output signal $y(s)$ and the control signal $u(s)$ is:
\begin{equation}
    G_\mathrm{UN}(s)=\frac{-C_\mathrm{FB}(s)}{1+G_\mathrm{OL}(s)}, 
    \label{eqn:noise_sensitivity_tf}
\end{equation}
which indicates that both considered control structures achieve the same level of noise sensitivity.

For oADRC, the transfer function from the reference signal $r(s)$ to the tracking error $e(s)$ is: 
\begin{equation}
    G_\mathrm{ER(o)}(s)=1-\frac{C_\mathrm{PF}(s)G_\mathrm{OL}(s)}{1+G_\mathrm{OL}(s)}, 
    \label{eqn:tracking_error_oADRC_tf}
\end{equation}
and for the eADRC is:
\begin{equation}
    G_\mathrm{ER(e)}(s)=1-\frac{G_\mathrm{OL}(s)}{1+G_\mathrm{OL}(s)}. 
    \label{eqn:tracking_error_eADRC_tf}
\end{equation}
In both above instances, there is a direct influence of $G_\mathrm{PF}(s)$. Therefore, this influence is further investigated. For the purpose of the investigation, two mathematical models of common industrial plants are introduced, namely a first-order system model ($n=1$) with time-delay:
\begin{equation}
    G_{\textrm{P1}}(s)=\frac{1}{s+1}e^{-0.2s}, 
    \label{eqn:first_order_plant}
\end{equation}
and a normalized second-order system model ($n=2$):
\begin{equation}
    G_{\textrm{P2}}(s)=\frac{1}{s^2+2s+1}, 
    \label{eqn:second_order_plant}
\end{equation}
The pre-filter and feedback transfer functions for the considered first- and second-order plant models are presented in \refTable{tab:First_order_transfer_function} and~\refTable{tab:Second_order_transfer_function}, respectively.

\begin{table*}
    \centering
    \caption{Transfer functions for first-order plant model ($n=1$).}
    \label{tab:First_order_transfer_function}
    {\renewcommand{\arraystretch}{2}
    \begin{tabular}{ccc}
    \hline
    \textbf{Control scheme} & \textbf{Pre-filter transfer function} & \textbf{Feedback transfer function}                                                           \\ \hline
    oADRC (case A)     & $G_\mathrm{PF}(s)= \frac{k_1\cdot (s^2+l_1s+l_2)}{(k_1l_1+l_2)\cdot s+k_1l_2}$  & $ G_\mathrm{FB}(s)=\frac{1}{b_0}\cdot \frac{(k_1l_1+l_2)\cdot s+k_1l_2}{s^2+(k_1+l_2) \cdot s}$ \\ 
    oADRC (case B) & $ G_\mathrm{PF}(s)= \frac{(k_1+s)\cdot(s^2+l_1s+l_2)}{(k_1l_1+l_2)\cdot s+k_1l_2}$ & $ G_\mathrm{FB}(s)=\frac{1}{b_0}\cdot \frac{(k_1l_1+l_2)\cdot s+k_1l_2}{s^2+(k_1+l_2)\cdot s} $     \\ 
    eADRC & \textbf{---}                  & $ G_\mathrm{FB}(s)=\frac{1}{b_0}\cdot \frac{(k_1l_1+l_2)\cdot s+k_1l_2}{s^2+(k_1+l_2)\cdot s}$  \\ \hline
    \end{tabular}%
    }
\end{table*}

\begin{table*}
    \centering
    \caption{Transfer functions for second-order plant model ($n=2$).}
    \label{tab:Second_order_transfer_function}
    {\renewcommand{\arraystretch}{2}
    \begin{tabular}{ccc}
    \hline
    \textbf{Control scheme} & \textbf{Pre-filter transfer function} & \textbf{Feedback transfer function}                                                           \\ \hline
    oADRC (case A)     &$ G_\mathrm{PF}(s)= \frac{k_1\cdot (s^3+l_1s^2+l_2s+l_3)}{(k_1l_1+k_2l_2+l_3)\cdot s^2+(k_1l_2+k_2l_3)\cdot s+k_1l_3}$   & $ G_\mathrm{FB}(s)=\frac{1}{b_0}\cdot \frac{(k_1l_1+k_2l_2+l_3)\cdot s^2+(k_1l_2+k_2l_3)\cdot s+k_1l_3}{s^3+(k_2+l_1)\cdot s^2+(l_2+k_1+l_1k_2)\cdot s}$  \\ 
    oADRC (case B) &  $G_\mathrm{PF}(s)= \frac{(s^2+k_2s+k_1)\cdot (s^3+l_1s^2+l_2s+l_3)}{(k_1l_1+k_2l_2+l_3)\cdot s^2+(k_1l_2+k_2l_3)\cdot s+k_1l_3}$& $ G_\mathrm{FB}(s)=\frac{1}{b_0}\cdot \frac{(k_1l_1+k_2l_2+l_3)\cdot s^2+(k_1l_2+k_2l_3)\cdot s+k_1l_3}{s^3+(k_2+l_1)\cdot s^2+(l_2+k_1+l_1k_2)\cdot s}$     \\ 
    eADRC & \textbf{---}                    & $ G_\mathrm{FB}(s)=\frac{1}{b_0}\cdot \frac{(k_1l_1+k_2l_2+l_3)\cdot s^2+(k_1l_2+k_2l_3)\cdot s+k_1l_3}{s^3+(k_2+l_1)\cdot s^2+(l_2+k_1+l_1k_2)\cdot s}$   \\ \hline
    \end{tabular}%
    }
\end{table*}

\begin{figure}[htb!]
     \centering
     \begin{subfigure}[b]{0.8\columnwidth}
         \centering
         \includegraphics[width=\columnwidth]{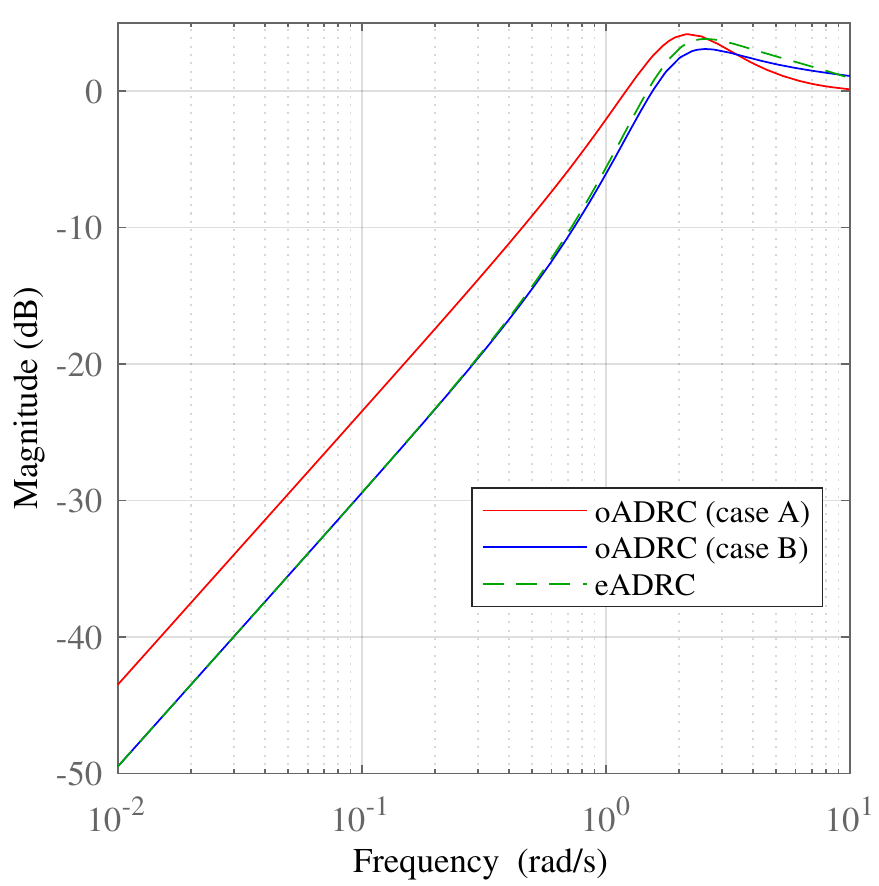}
         \caption{}
         \label{fig:Tracking_performance_ADRC_eADRC_Gp1}
     \end{subfigure}
     \begin{subfigure}[b]{0.8\columnwidth}
         \centering
         \includegraphics[width=\columnwidth]{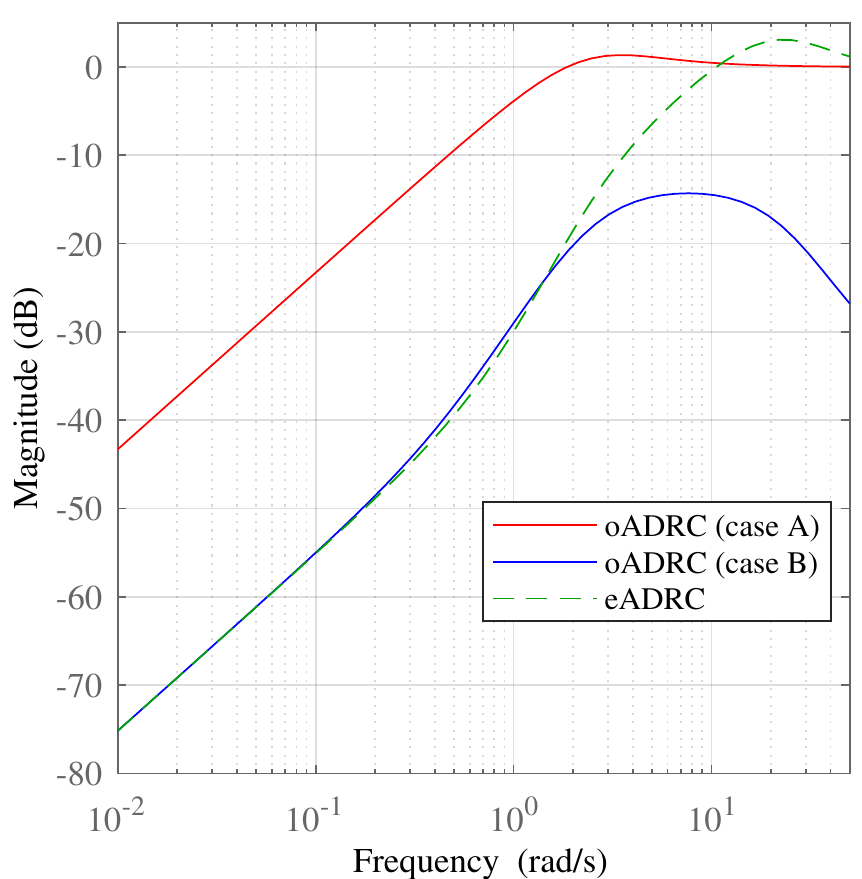}
         \caption{}
         \label{fig:Tracking_performance_ADRC_eADRC_Gp2}
     \end{subfigure}
        \caption{Comparative analyses of the reference tracking performances of oADRC (cases A and B) and eADRC for plant model $G_{\textrm{P1}}(s)$ (a) and $G_{\textrm{P2}}(s)$ (b).}
        \label{fig:Tracking_performance_ADRC_eADRC}
\end{figure}

The comparative frequency responses of~\refEq{eqn:tracking_error_oADRC_tf} (for cases A and B) and \refEq{eqn:tracking_error_eADRC_tf}, with design parameters $\omega_{\textrm{CL}}=3$rad/s and $k_\mathrm{ESO}=8$, are shown in \refFig{fig:Tracking_performance_ADRC_eADRC_Gp1} for the first-order plant~\refEq{eqn:first_order_plant} and in \refFig{fig:Tracking_performance_ADRC_eADRC_Gp2} for the second-order plant~\refEq{eqn:second_order_plant}.

\begin{figure*}
     \centering
     \begin{subfigure}[b]{0.9\textwidth}
         \centering
         \includegraphics[width=0.9\textwidth]{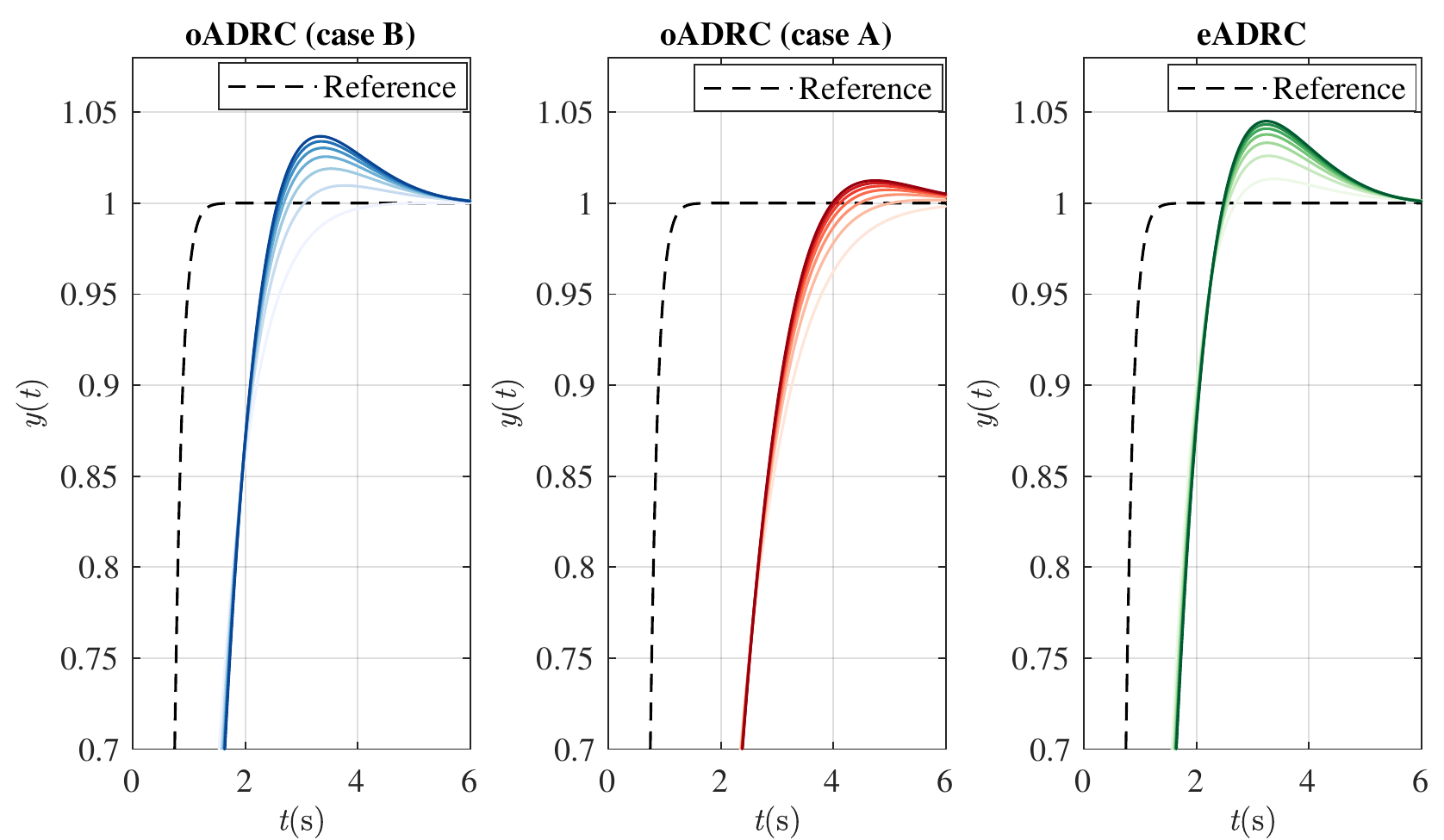}
         \caption{System response (zoomed figure) for a reference signal (dashed line) using fixed $\omega_\mathrm{CL}=1.5$rad/s and linearly increasing values of $k_\mathrm{ESO}$, from $k_\mathrm{ESO}=6$ (low color saturation) to $k_\mathrm{ESO}=24$ (high color saturation).} 
         \label{fig:ParamVarKesoGp1}
     \end{subfigure}
     \hfill
     \begin{subfigure}[b]{0.9\textwidth}
         \centering
         \includegraphics[width=0.9\textwidth]{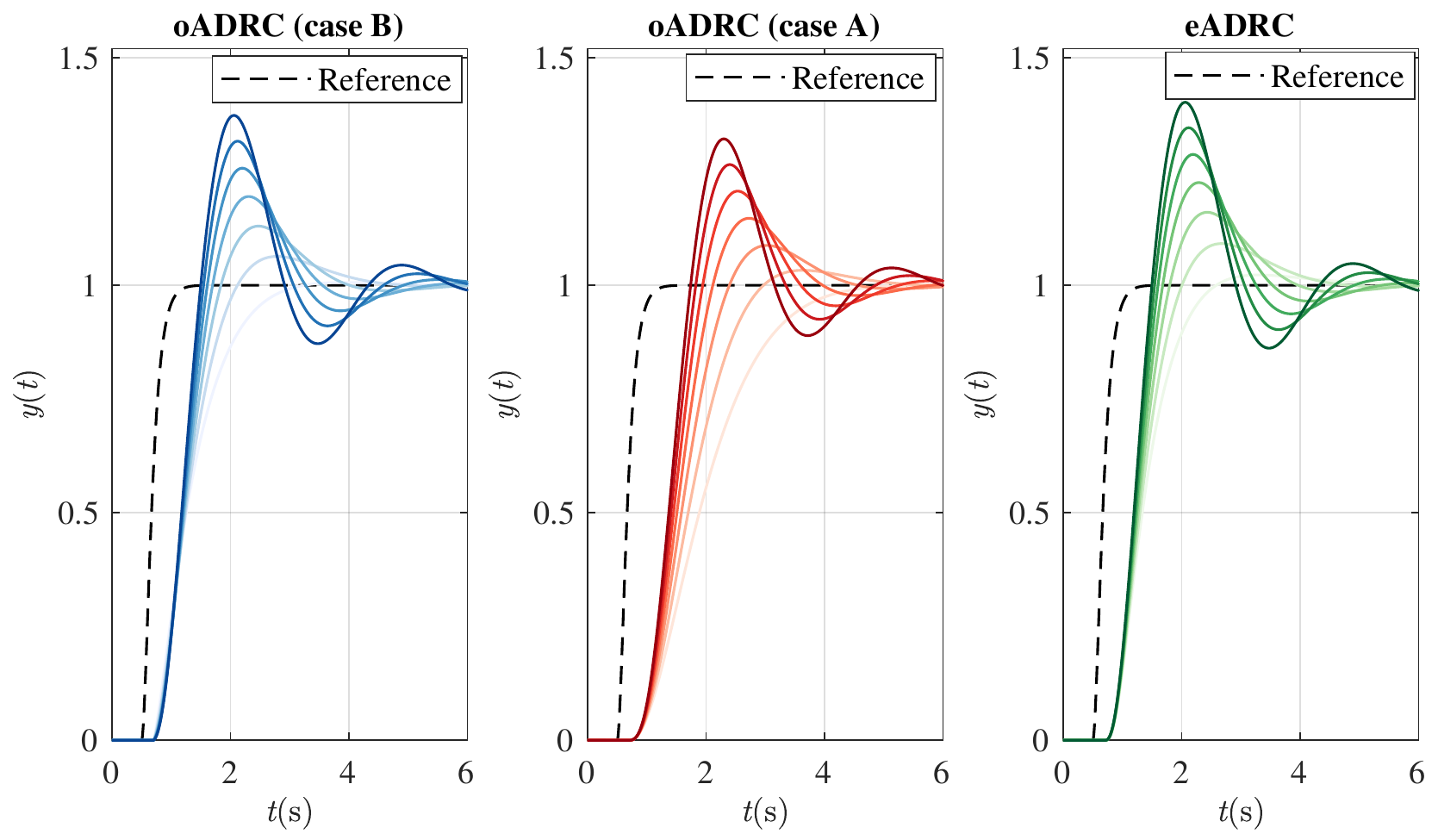}
         \caption{System response for a reference signal (dashed line) using fixed $k_\mathrm{ESO}=8$ and linearly increasing values of $\omega_\mathrm{CL}$, from $\omega_\mathrm{CL}=1.5$ rad/s (low color saturation) to $\omega_\mathrm{CL}=4.5$ rad/s (high color saturation).}
         \label{fig:ParamVarOmegaGp1}
     \end{subfigure}
        \caption{Influence of tuning parameters on the tracking performance of system $G_{\textrm{P1}}(s)$ with the control structures oADRC in case~B (left), oADRC in case~A (middle), and eADRC (right).}
        \label{fig:ParamVarGp1}
\end{figure*}

\begin{figure*}
     \centering
     \begin{subfigure}[b]{0.9\textwidth}
         \centering
         \includegraphics[width=0.9\textwidth]{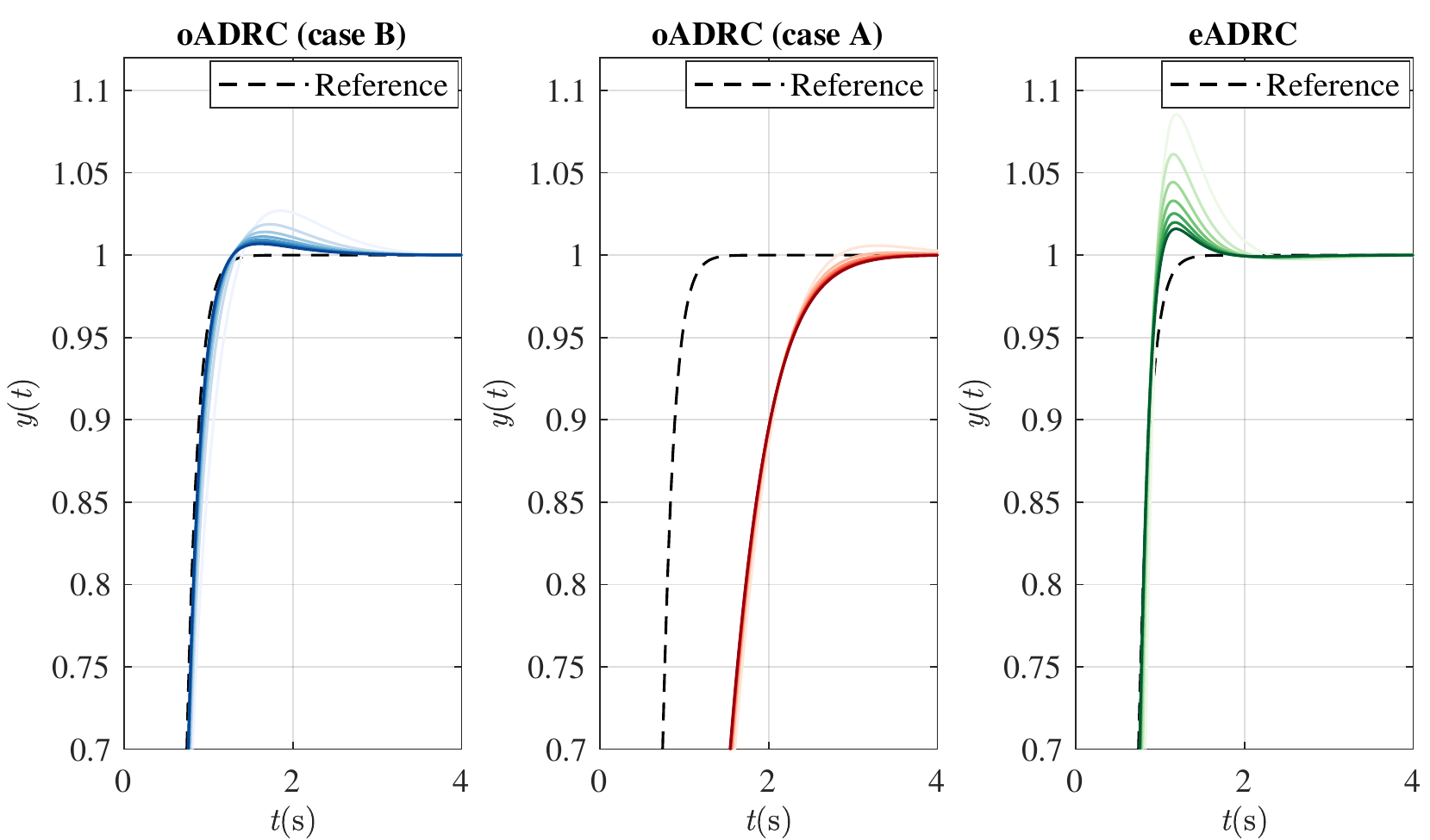}
         \caption{System response (zoomed figure) for a reference signal (dashed line) using fixed $\omega_\mathrm{CL}=3$rad/s and linearly increasing values of $k_\mathrm{ESO}$, from $k_\mathrm{ESO}=6$ (low color saturation) to $k_\mathrm{ESO}=24$ (high color saturation).}
         \label{fig:ParamVarKesoGp2}
     \end{subfigure}
     \hfill
     \begin{subfigure}[b]{0.9\textwidth}
         \centering
         \includegraphics[width=0.9\textwidth]{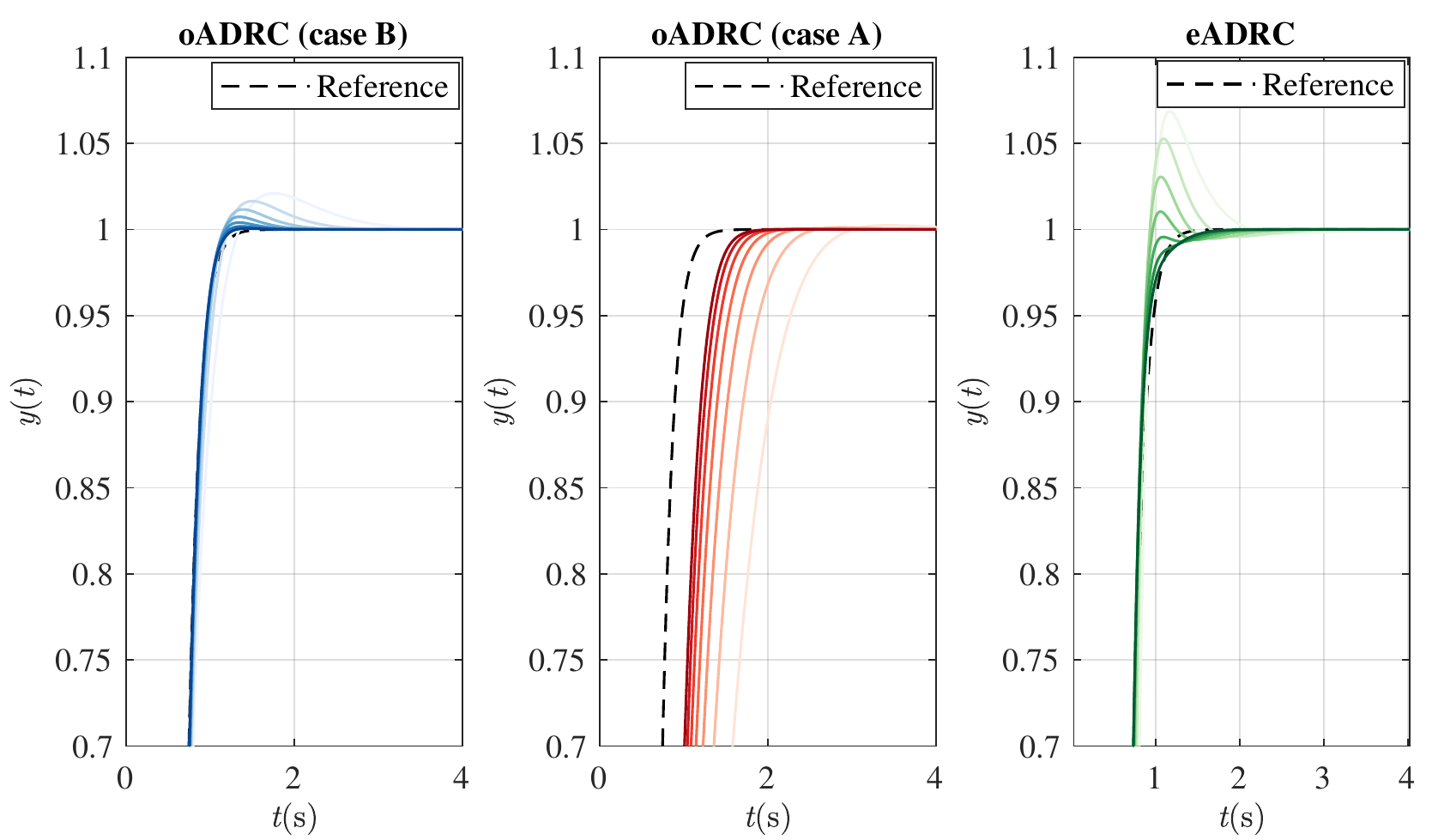}
         \caption{System response for a reference signal (dashed line) using fixed $k_\mathrm{ESO}=8$ and linearly increasing values of $\omega_\mathrm{CL}$, from $\omega_\mathrm{CL}=3$ rad/s (low color saturation) to $\omega_\mathrm{CL}=9$ rad/s (high color saturation).}
         \label{fig:ParamVarOmegaGp2}
     \end{subfigure}
        \caption{Influence of tuning parameters on the tracking performance of system $G_{\textrm{P2}}(s)$ with the control structures oADRC in case~B (left), oADRC in case~A (middle), and eADRC (right).}
        \label{fig:ParamVarGp2}
\end{figure*}

The obtained results from~\refFig{fig:Tracking_performance_ADRC_eADRC_Gp1} and~\refFig{fig:Tracking_performance_ADRC_eADRC_Gp2} show that the oADRC with the reference derivatives included in the control law (case~B) achieves significantly lower tracking error than the oADRC without reference derivatives in the control law (case~A) over the entire reference frequency range of interest (lower than the tuned system closed-loop bandwidth $\omega_\mathrm{CL}=3$rad/s). Also, one can notice that the eADRC provides practically the same tracking performance as the oADRC (case~B), but without requiring the availability of the reference derivatives in the control law. It can be explained by the fact that eADRC indirectly estimates the reference derivatives by the ESO \refEq{eqn:eADRC_CT_SS_Observer} and use them in the controller synthesis, which can be seen in the definition of the total disturbance term $f_{\textrm{e}}(t)$ under~\refEq{eqn:eADRC_representation_of_general_nth_order_plant}.  

\paragraph{Transient Performance} The transient performances of the considered ADRC structures are analyzed in time-domain using numerical simulations with the plant models \refEq{eqn:first_order_plant} and  \refEq{eqn:second_order_plant}. 
First, the simulations are conducted with the reference signal in form of a step signal filtered by a low-pass filter $G_{\textrm{F}}(s)=\frac{1}{(0.1s+1)^2}$ and with various tuning parameters, in two scenarios: with fixed $\omega_\mathrm{CL}$ and variable $k_\mathrm{ESO}$, and with fixed $k_\mathrm{ESO}$ and variable $\omega_\mathrm{CL}$. The obtained results are gathered in~\refFig{fig:ParamVarGp1} and~\refFig{fig:ParamVarGp2}, respectively.

From the obtained results, one can notice that in all the simulated scenarios, both oADRC structures gave responses with lower overshoots than eADRC, especially for the second-order system $G_{\textrm{P2}}(s)$, where oADRC structures provide practically minimal overshoot. Therefore, it can be concluded from the performed tests that the oADRC schemes, thanks to the presence of component $G_{\textrm{PF}}(s)$, represent a better options than eADRC in terms the ability to shape the system response during the transient stage. Also, one can conclude that the oADRC (case~B) and eADRC enable faster responses compared to oADRC (case~A) due to the use of the reference signal derivatives in the control law, which are interpreted as actionable information that is consciously used by the control designer to improve the control performance.

Additionally, comparing the responses of the first- and second-order plants for all the considered controller structures, one can notice that in the instance of second-order system $G_{\textrm{P2}}(s)$, increasing parameters $k_\textrm{ESO}$ and $\omega_\mathrm{CL}$ leads to smaller overshoot. That is due to higher observer and controller dynamics, which implies faster convergence of the total disturbance estimation and better tracking performance. On the other hand, in the instance of first-order system $G_{\textrm{P1}}(s)$, increasing parameters $k_\textrm{ESO}$ and $\omega_\mathrm{CL}$ lead to larger overshoot. That is not surprising as it is a consequence of the presence of time-delay in the plant model.     

\begin{figure}[p]
     \centering
     \begin{subfigure}[b]{.9\columnwidth}
         \centering
         \includegraphics[width=\textwidth]{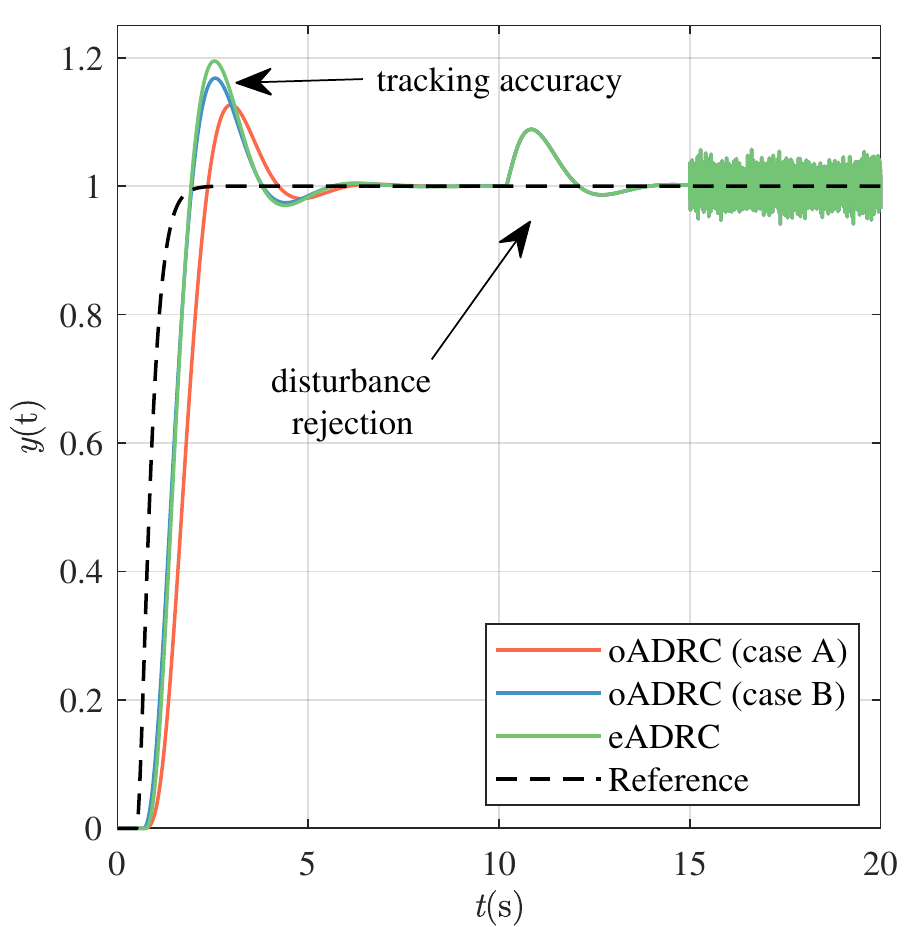}
         \caption{}
         \label{fig:Time_domain_step_response_Gp1}
     \end{subfigure}
     \begin{subfigure}[b]{.9\columnwidth}
         \centering
         \includegraphics[width=\textwidth]{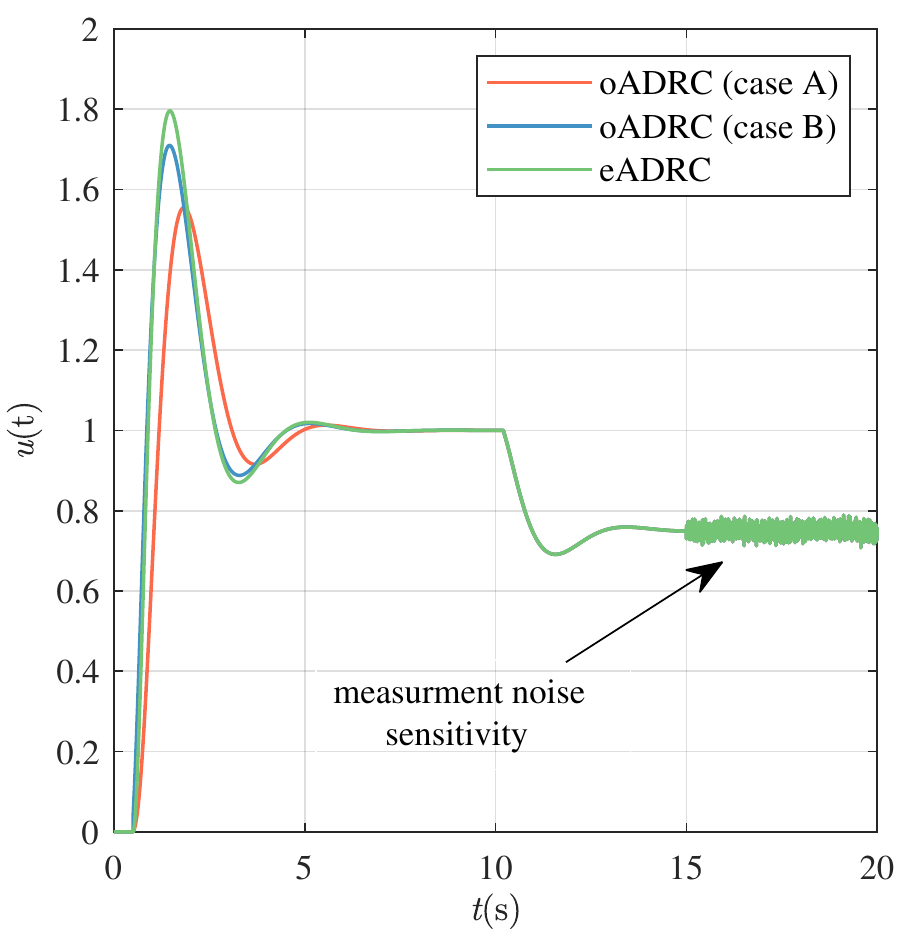}
         \caption{}
         \label{fig:Time_domain_step_control_signal_Gp1}
     \end{subfigure}
        \caption{Comparative analyses of transient performances of oADRC (cases A and B) and eADRC for plant model $G_{\textrm{P1}}(s)$, where (a) is the system output and (b) is the control signal. Note: during the disturbance rejection test ($10\textrm{s} \leq t < 15\textrm{s}$) and measurement noise sensitivity test ($15\textrm{s} \leq t \leq 20\textrm{s}$), the signals overlap. Note: remember that $G_{\textrm{P1}}(s)$ has time-delay, which especially influences the initial response (overshoot).}
        \label{fig:Time_domain_analyses_ADRC_eADRC_Gp1}
\end{figure}

\begin{figure}[p]
     \centering
     \begin{subfigure}[b]{.9\columnwidth}
         \centering
         \includegraphics[width=\textwidth]{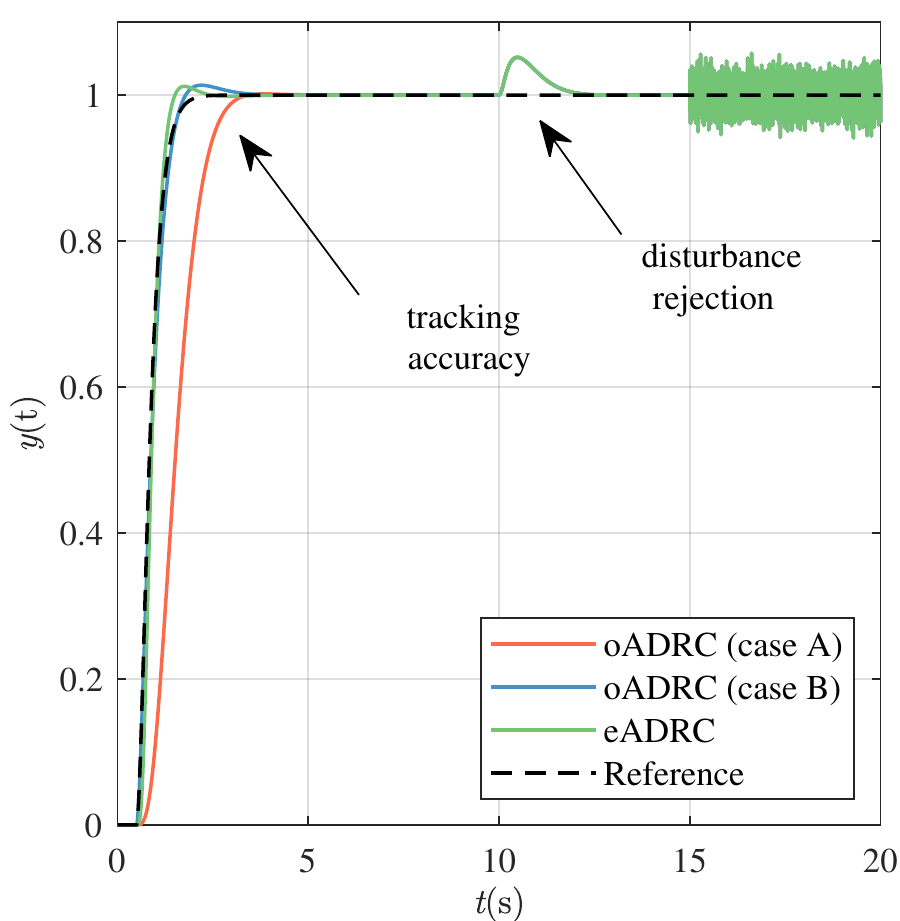}
         \caption{}
         \label{fig:Time_domain_step_response_Gp2}
     \end{subfigure}
     \begin{subfigure}[b]{.9\columnwidth}
         \centering
         \includegraphics[width=\textwidth]{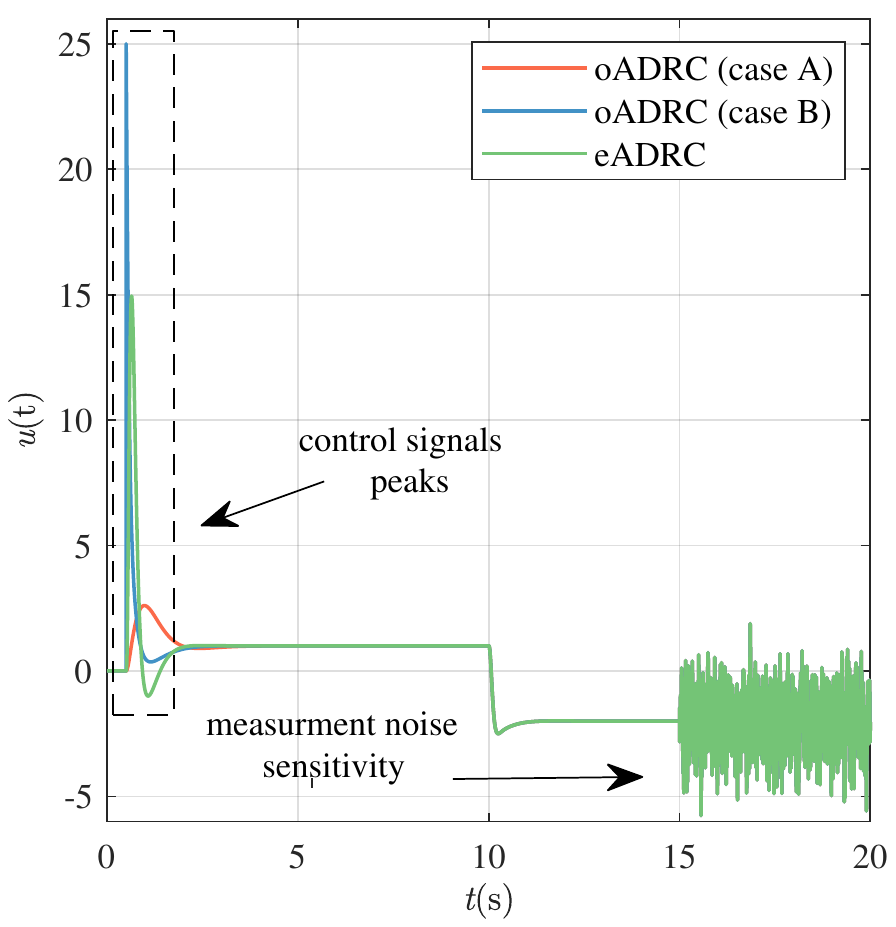}
         \caption{}
         \label{fig:Time_domain_step_control_signal_Gp2}
     \end{subfigure}
        \caption{Comparative analyses of transient performances of oADRC (cases A and B) and eADRC for plant model $G_{\textrm{P2}}(s)$, where (a) is the system output and (b) is the control signal. Note: during the disturbance rejection test ($10\textrm{s} \leq t < 15\textrm{s}$) and measurement noise sensitivity test ($15\textrm{s} \leq t \leq 20\textrm{s}$), the signals overlap.}
        \label{fig:Time_domain_analyses_ADRC_eADRC_Gp2}
\end{figure}

In order to compare energy requirements of the considered controllers as well as to show time-domain disturbance rejection performance and measurement noise sensitivity, the simulation with reference signal in form of a step signal filtered by a low-pass filter $G_{\textrm{F}}(s)=\frac{1}{(0.2s+1)^2}$ in the presence of step disturbance from $t \geq 10$s and measurement noise from $t \geq 15$s, are performed. The obtained system responses for plants \refEq{eqn:first_order_plant} and  \refEq{eqn:first_order_plant} are depicted in~\refFig{fig:Time_domain_analyses_ADRC_eADRC_Gp1} and~\refFig{fig:Time_domain_analyses_ADRC_eADRC_Gp2}, respectively. The simulations are performed for fixed tuning parameters $\omega_{\textrm{CL}}=3$rad/s and $k_\mathrm{ESO}=8$. Please note that the overshoot effect in \refFig{fig:Time_domain_analyses_ADRC_eADRC_Gp1} is also closely related with time-delay in $G_{\textrm{P1}}(s)$, which additionally limits the upper range of a practically feasible desired closed-loop bandwidth.


From the figures, it can be noted that the price for the faster responses of oADRC (case~B) and eADRC is the increased level of control signal, especially in the second-order system simulation with oADRC in \refFig{fig:Time_domain_step_control_signal_Gp2}. It is a consequence of a relatively large value of the reference signal's second time-derivative, which is necessary for the synthesis of the control law. Therefore, the application of the ADRC (case~B) structure for higher-order systems ($n>2$) can be limited in practice, due to a potentially large value of the reference $n$-th derivatives. 
Regarding the disturbance rejection performances, as expected, it is evident that all the considered structures have the same disturbance rejection responses, which confirms the previously made conclusion resulting from~\refEq{eqn:dist_rejection_tf}. Similarly, looking at the control signals at time $t \geq 15$s in~\refFig{fig:Time_domain_step_control_signal_Gp1} and~\refFig{fig:Time_domain_step_control_signal_Gp2}), one can notice identical levels of measurement noise sensitivity, which confirms conclusions made in $s$-domain in~\refEq{eqn:noise_sensitivity_tf}.

\textbf{Answer}. To address the question posed in the title of this subsection, the pre-filter transfer function $G_\mathrm{PF}(s)$ has, based on the conducted tests, impact on the tracking performance during the transient control stage. The analysis has shown that in steady-state, $G_\mathrm{PF}(s)$ has no effect on the stability margins, sensitivity to measurement noise, and disturbance rejection, as well as a negligible impact on the steady-state reference tracking performance when it is realized by including reference derivative terms (oADRC in case~B). The inclusion of time-delay in $G_{\textrm{P1}}(s)$ was deliberate, as this type of system is often seen in industrial practice, but it does not change the findings or the general conclusions from this section.

\subsection{Are the Considered ADRC Transfer Functions Realizable?}
\label{sec:Q3realizability}

To enable realizability of the considered transfer function representations of oADRC and eADRC, it is necessary to make all the transfer functions within those structures realizable.

For eADRC, it requires only the feedback transfer function \refEq{eqn:Feedback_TF} for its implementation (cf. \refFig{fig:ADRC_eADRC_tf_forms}), hence the realizability of this transfer function has to be checked. By analyzing~\refEq{eqn:Feedback_TF}, one can see that its numerator and denominator orders depend on matrix $\Matrix{A}_{\textrm{CL}}$. The fact that the last row and rightmost column of the $\Matrix{A}_{\textrm{CL}}$ are zero, and by assuming non-zero observer and controller gains, makes its rank equal to $n$. At the same time, the order of numerator polynomial of $G_{\mathrm{FB}}(s)$ is $n$, while its denominator polynomial is $n+1$, which is also evident from \refTable{tab:First_order_transfer_function} and~\refTable{tab:Second_order_transfer_function}. It can be thus concluded that the denominator polynomial order exceeds the numerator polynomial order, meaning that $G_{\mathrm{FB}}(s)$ (and eADRC as a result) is a strictly proper and a realizable transfer function.

For oADRC, its realizability depends on the realizability of \refEq{eqn:PreFilter_TF_with_ref_derivatives_reformulation}. Assuming that $G_{\mathrm{R}}(s)$ can be realized as the sum of $n+1$ inputs (i.e. reference signal and its $n$ derivatives), multiplied by the corresponding coefficients, the problem of realizability is focused on  \refEq{eqn:PreFilter_TF_without_ref_derivatives}. By its inspection, one can see that due to $A_{\textrm{CL}}$ having rank $n$, \refEq{eqn:PreFilter_TF_without_ref_derivatives}  is not proper, meaning the transfer function is not realizable, because its numerator polynomial has order $n+1$ but its denominator polynomial has order $n$ (cf. \refTable{tab:First_order_transfer_function} and \refTable{tab:Second_order_transfer_function}). Therefore, oADRC cannot be realized in the form as presented in \refFig{fig:ADRC_eADRC_tf_forms}.

\begin{figure*}[t!]
    \centering
    \includegraphics[width=.8\textwidth]{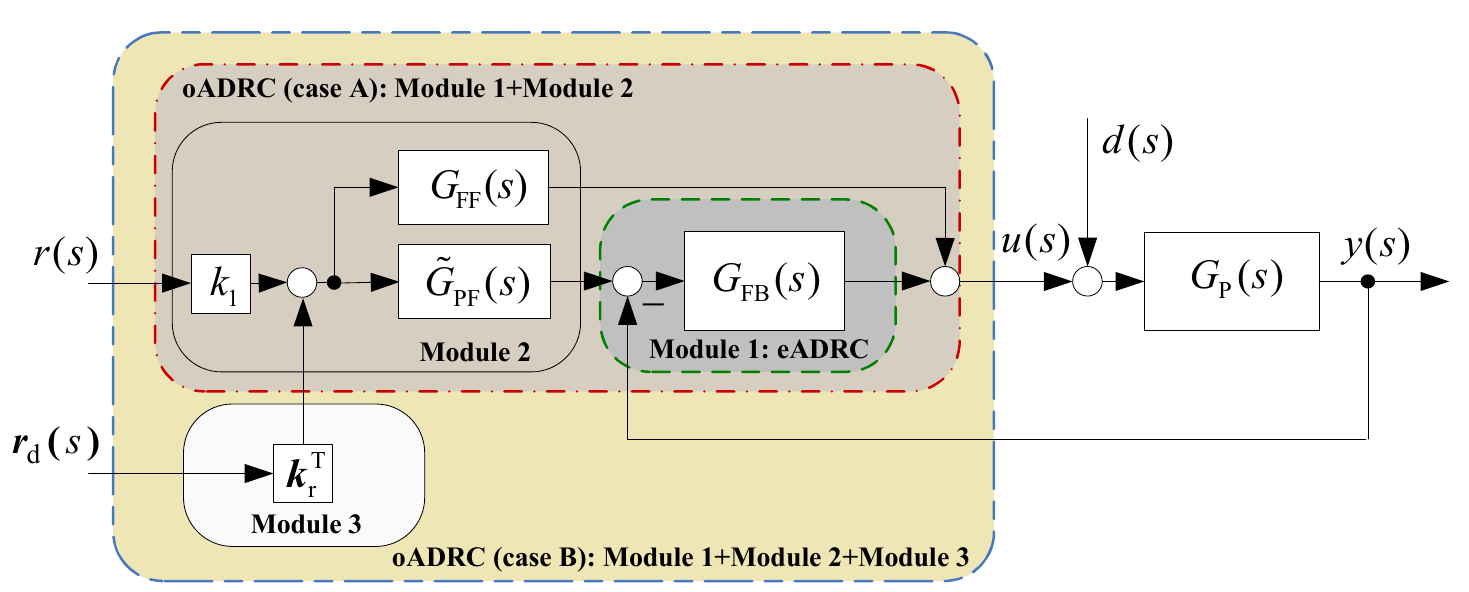}
    \caption{Proposed modular approach to designing customized ADRC-based structures in form of realizable transfer functions, with visual cues for obtaining nominal oADRC case A (dashed-dotted red), oADRC case B (short-long dashed blue), and eADRC (dashed green). Interpretation of the modules: Module~1---eADRC, which is the \textit{bare-bones} version of ADRC consisting just the feedback transfer function $G_{\textrm{FB}}(s)$; Module~2---eADRC is transformed into oADRC by adding a reference channel with a prefilter $\tilde{G}_{\textrm{PF}}(s)$ (for shaping transient performance) and a feedforward term $G_{\textrm{FF}}(s)$ (for making the transfer function realizable); Module~3---it additionally equips oADRC with the derivatives of the reference signal (assuming they are available) to improve the tracking performance (especially helpful in motion control tasks).}
     \label{fig:Modular_realization_structure}
 \end{figure*}

Following methodology presented in~\cite{HerbstTF}, this problem can be solved by a reformulation of the transfer function representation of oADRC from the standard pre-filter expression \refEq{eqn:ADRC_CT_TF_Representation} to the combination of pre-filter and feedforward:
\begin{align}
    u(s) &= G_\mathrm{FB}(s)\cdot\left[G_\mathrm{R}(s)\tilde{G}_\mathrm{PF}(s)r(s)-y(s)\right]+\notag\\
    &+ G_\mathrm{R}(s)G_\mathrm{FF}(s)r(s),
    \label{eqn:ADRC_CT_TF_Representation_feedforward_realizable}
 \end{align}  
with the introduced feedforward transfer function, which carries the $s^{n+1}$ numerator term of \refEq{eqn:PreFilter_TF_without_ref_derivatives}:
\begin{equation}
    G_\mathrm{FF}(s)=\frac{s^{n+1}}{b_0 \cdot \operatorname{det}\left( s \Matrix{I} - \Matrix{A}_\mathrm{CL} \right)},
    \label{eqn:ADRC_feedforward_TF}
 \end{equation}   
and modified pre-filter transfer function:
\begin{gather}
\tilde{G}_{\mathrm{PF}}(s)   =
    \frac{\operatorname{det}( s \Matrix{I} - \Matrix{A}_\mathrm{CL}) -s^{n+1}-\frac{1}{b_0}\begin{pmatrix}
        k^\Transpose & 1
    \end{pmatrix} \operatorname{adj}\left( s \Matrix{I} - \Matrix{A}_\mathrm{CL} \right)\cdot\Vector{b}}{\begin{pmatrix}
        k^\Transpose & 1
    \end{pmatrix} \operatorname{adj}\left( s \Matrix{I} - \Matrix{A}_\mathrm{CL} \right)\cdot\Vector{l}}.
   \label{eqn:PreFilter_TF_modified}
\end{gather}
One can note that order of numerator and denominator of \refEq{eqn:ADRC_feedforward_TF} will be $n$ due to cancellation of its numerator with the integrator introduced by $\operatorname{det}( s \Matrix{I} - \Matrix{A}_\mathrm{CL})$, which means that this transfer function can be realized as $n$-th order high-pass filter. On the other hand, modified pre-filter transfer function \refEq{eqn:PreFilter_TF_modified} has $n$-th order denominator, same as \refEq{eqn:PreFilter_TF_with_ref_derivatives}, and $n$-th order numerator, what makes it proper and realizable. An additional advantage of this structure is that integrator component exists only in the feedback transfer function, which is important from the practical realization point of view \cite{HerbstTF}.  

The graphical representation of the proposed realizable forms of the considered oADRC (in cases A and B) and eADRC, based on a modular approach, are shown in \refFig{fig:Modular_realization_structure}. One can see that eADRC is the least complex structure which requires only one module ('Module 1') for its realization. The transition from eADRC to oADRC (case~A) is possible by adding 'Module 2'. The oADRC (case~B) represents the most complex structure.


\textbf{Answer}. To address the question posed in the title of this subsection, the transfer function representation of the eADRC scheme has a realizable form and thus can be directly used. For oADRC, we have shown a simple modification to its nominal form that allows to represent the algorithm in an realizable way by utilizing the combination of pre-filter and feedforward concepts.  

\section{Summary}
\label{sec:Summary}

In this paper, output- and error-based versions of the ADRC scheme were investigated. Through an analysis in $s$-domain, a condition of equivalence between the two structures has been established. It was found that the only difference between the two structures is the presence of a so-called pre-filter transfer function ($G_\textrm{PF}(s)$). The frequency domain analysis has shown that in steady-state, $G_\textrm{PF}(s)$ has no effect on the stability margins, sensitivity to measurement noise, or disturbance rejection, but has impact on the  steady-state reference tracking performance. It was also deduced that synthesising and utilizing $G_\textrm{PF}(s)$ without reference derivatives (i.e. oADRC in case~A) reduces steady-state reference tracking performance, whereas adding reference derivatives in the oADRC (i.e. case~B) provides quantitatively better performance, practically the same as in the eADRC control structure (where $G_\textrm{PF}(s)=1$ is assumed). Additionally, the conducted investigation of the influence of $G_\textrm{PF}(s)$ on basic control performance criteria in time-domain has revealed that the ADRC with the pre-filter allows the control designer to better shape the tracking performance in the transient stage, which can be beneficial in certain practical scenarios. 

\begin{figure*}[t]
    \centering
    \includegraphics[width=0.75\textwidth]{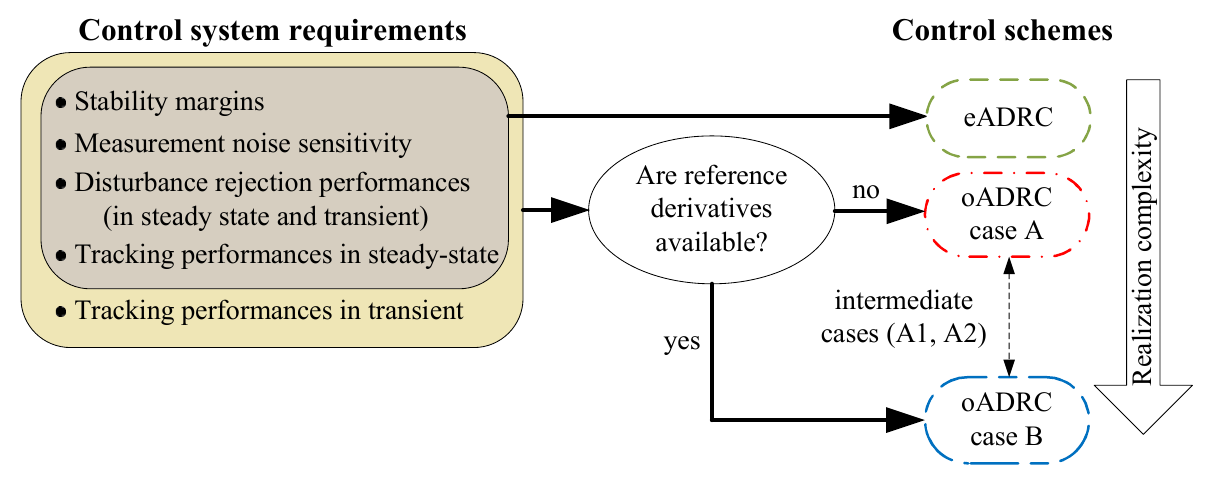}
    \caption{Proposed heuristic decision tree for choosing ADRC structure (for the interpretation of 'intermediate cases (A1, A2)' --- see \refRem{lab:CaseRefDerivatives} and the Appendix).}
     \label{fig:Decision_tree_ADRC}
\end{figure*}

That is why, in order to assist the control designers in choosing most suitable ADRC-based solution for their applications, a summary of the conducted systematic comparison has been graphically represented with a (heuristic) decision tree, depicted in \refFig{fig:Decision_tree_ADRC}. In short, it can be seen that the eADRC has a simpler, more compact design, resulting in a simpler implementation (due to the lack of the pre-filter), but for the price of reducing the control over transient tracking performance. The eADRC also gives more freedom to the control designer to construct custom control solutions since the pre-filter $G_\textrm{PF}(s)$ is not built-in (imposed) in the design in advance (as it is the case in oADRC). Because there are different control scenarios that call for a different control solution, a modular approach has been chosen in this work. Certain modules have been identified (see~\refFig{fig:Modular_realization_structure}) that aid the process of designing the desired ADRC solution. 

Finally, in order to increase the potential applicability of both considered ADRC schemes, their realizable transfer functions forms were given.

\section*{Appendix --- Influence of Reference Derivatives}
\label{sec:appendixA}


In order to show the influence of various levels of knowledge about the reference derivatives on the tracking quality in oADRC, we introduce a third-order ($n=3$) plant model: 
\begin{equation}
G_{\mathrm{P3}}(s)=\frac{1}{s^3+s^2+s+1},
   \label{eqn:Third_order_plant_model}
\end{equation}
that does not represent a physical plant and is used here solely as a toy example which happens to be useful to us in supporting the idea we want to convey here.

\begin{figure*}[t]
    \centering
    \includegraphics[width=.8\textwidth]{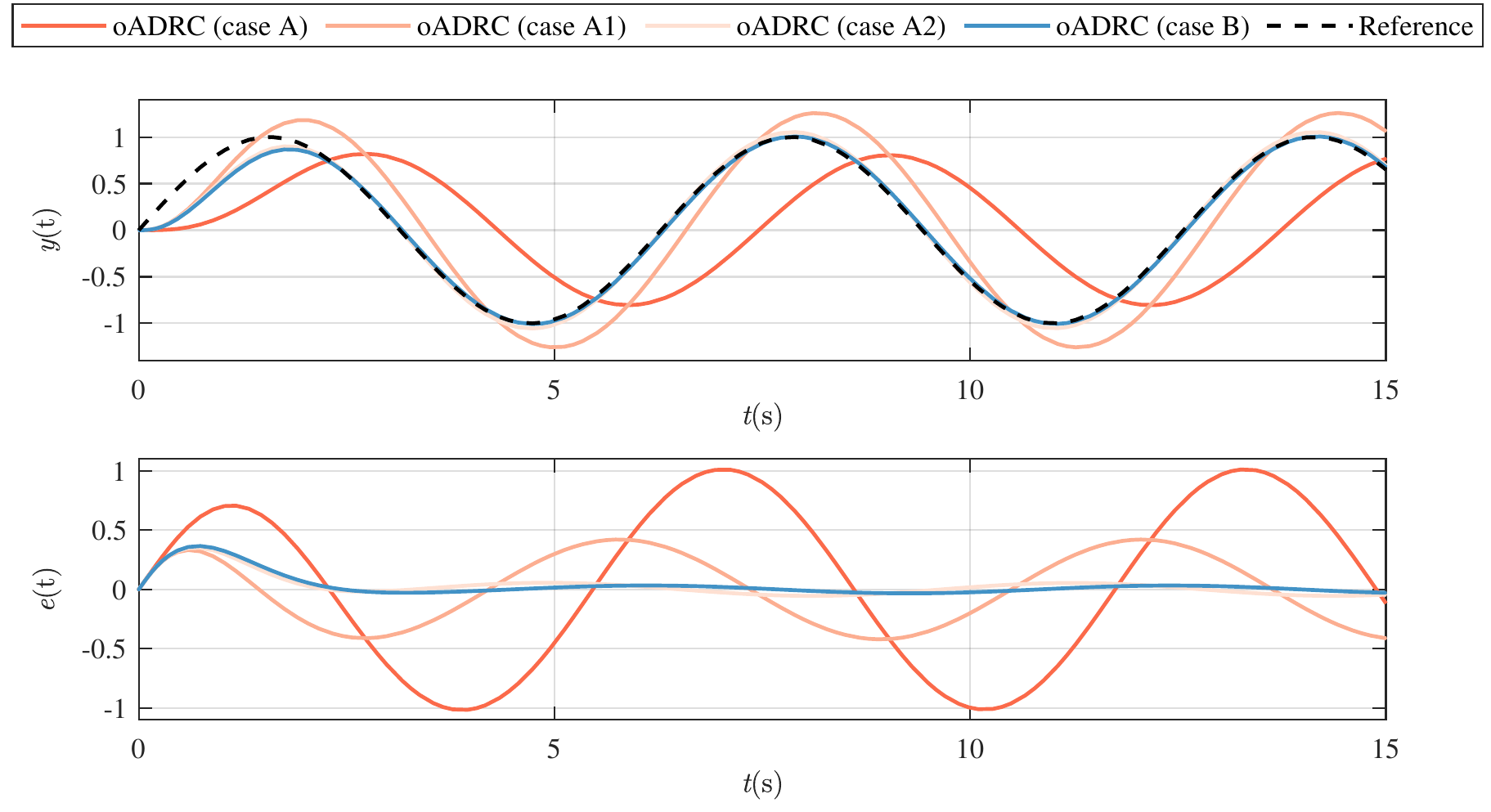}
    \caption{Reference tracking performances of oADRC in cases A and B, as well as two intermediate cases A1 and A2, with reference signal and system outputs (top) and tracking error (bottom).}
     \label{fig:oADRC_for_3th_order_plant}
     
\end{figure*}

Next, a numerical simulation is performed with oADRC controllers (in previously defined cases A and B) as well as two intermediate cases, which include terms related to the first and second reference derivative in~\refEq{eqn:ADRC_CT_SS_Controller}. Those intermediate cases of oADRC are denoted as A1 and A2 and are formally defined in~\refTable{tab:ConsideredControlSchemesEXT}. The controllers considered in the simulation are all tuned using the 'bandwidth parameterization' approach with the same $\omega_\mathrm{CL}=2.5$rad/s and $k_\mathrm{ESO}=5$. The tracking performances obtained for a reference sinusoidal signal $r(t)=\sin(t)$ are shown in~\refFig{fig:oADRC_for_3th_order_plant}.

From \refFig{fig:oADRC_for_3th_order_plant}, it is evident that, by including more derivatives of the reference signal to the oADRC control law, the reference tracking performance is improved. This makes oADRC (case B) variant especially suitable for tracking control of dynamic signals. However, it should also be noted that adding more reference derivatives during the controller synthesis also makes the final controller more complex and that in industrial practice the reference derivatives are not always available. Therefore, the bottom line is that the selection of a specific ADRC controller structure should be done on a case-by-case basis.

\begin{table}[htb!]
\centering
\caption{Considered oADRC cases for $G_{\mathrm{P3}}(s)$ (cf.~\refTable{tab:ConsideredControlSchemes}).}
\label{tab:ConsideredControlSchemesEXT}
{\renewcommand{\arraystretch}{1.5}
\begin{tabular}{ccc}
\hline
\textbf{oADRC} & \textbf{Gains}  & \textbf{Assumption}                                                         \\ \hline
case A & $\Vector{k_\textrm{r}}^\Transpose =
    \begin{pmatrix}
        0 & 0 & 0
    \end{pmatrix}$ & $\dot{r},\ddot{r},r^{(3)}$ not available    \\ 
case A1 & $\Vector{k_\textrm{r}}^\Transpose =
    \begin{pmatrix}
        k_2 & 0 & 0
    \end{pmatrix}$ & only $\dot{r}$ available     \\ 
case A2 & $\Vector{k_\textrm{r}}^\Transpose =
    \begin{pmatrix}
        k_2 & k_3 & 0
    \end{pmatrix}$ & only $\dot{r},\ddot{r}$ available    \\ 
case B & $\Vector{k_\textrm{r}}^\Transpose =
    \begin{pmatrix}
        k_2 & k_3 & 1
    \end{pmatrix}$ & $\dot{r},\ddot{r},r^{(3)}$ available \\ \hline
\end{tabular}%
}
\end{table}

\section*{Statements and Declarations}

\begin{itemize}  
    \item R.~Madonski was supported by the Fundamental Research Funds for the Central Universities (project no. 21620335). 
    \item M.~Stankovic was supported by the International Foreign Expert Project Fund of Jinan University (project no. G2021199027L, coordinator: Dr.~Hui Deng).
    \item The authors have no competing interests to declare that are relevant to the content of this article.
\end{itemize}



\bibliographystyle{IEEEtran}
\bibliography{bibliography}

\end{document}